\documentclass[aip,jcp,reprint]{revtex4-1}
\usepackage{bm}
\usepackage{amsmath}
\usepackage{graphicx}
\usepackage{amssymb}
\usepackage[normalem]{ulem}
\usepackage[usenames]{color}
\begin{document}

\newcommand{\C}{\mathcal{C}}
\newcommand{\ee}{\mathrm{e}}
\newcommand{\eb}{\varepsilon_\mathrm{b}}
\newcommand{\ebt}{\tilde\varepsilon_\mathrm{b}}
\newcommand{\tobs}{t_\mathrm{obs}}
\newcommand{\nm}{n_\mathrm{max}}
\newcommand{\numb}{n_\mathrm{umb}}

\newcommand{\sub}[1]{ _{\mathrm{#1}}}
\newcommand{\LJ}[1]{ \mathcal{L}_{#1} }
\newcommand{\ecc}{\eb}
\newcommand{\kbt}{k_\mathrm{B} T}
\newcommand{\kb}{k_\mathrm{B}}
\newcommand{\kt}{k_\mathrm{B}T}
\newcommand{\Nc}{N_\mathrm{C}}
\newcommand{\rsub}[1]{_\mathrm{#1}}
\newcommand{\sbb}{s_\mathrm{b}}
\newcommand{\css}{c_\mathrm{ss}}
\newcommand{\gb}{g_\mathrm{b}}
\newcommand{\sconf}{\Delta s_\mathrm{c}}  

\newcommand{\bad}{\langle \Delta B(n)\rangle}  

\title{Mechanisms of kinetic trapping in self-assembly and phase transformation}
\author{Michael F. Hagan}
\affiliation{Department of Physics, Brandeis University, Waltham, MA}
\author{Oren M. Elrad}
\affiliation{Department of Physics, Brandeis University, Waltham, MA}
\author{Robert L. Jack}
\affiliation{Department of Physics, University of Bath, Bath BA2 7AY, United Kingdom}

\begin{abstract}
In self-assembly processes, kinetic trapping effects often hinder the formation of thermodynamically stable ordered states.  In a model of viral capsid assembly and in the phase transformation of a lattice gas, we show how simulations in a self-assembling steady state can be used to identify two distinct mechanisms of kinetic trapping.  We argue that one of these mechanisms can be adequately captured by kinetic rate equations, while the other involves a breakdown of theories that rely on cluster size as a reaction coordinate. We discuss how these observations might be useful in designing and optimising self-assembly reactions.
\end{abstract}

\maketitle

\section{Introduction}
\label{sec:intro}

In self-assembly processes, simple components come together spontaneously to form ordered products. Such processes abound in biology, where the ordered structures might be the outer shells of viruses ~\cite{expt_virus_misc,theory_virus_misc,Hagan2006,Jack2007}, extended one-dimensional filaments that make up the cytoskeleton~\cite{Yang2010a}, or ordered arrays of proteins on the surface of bacteria~\cite{Whitelam2010}.
In other areas of nano-science, self-assembled nanostructures made from customised DNA oligomers are being used to build ever more complex structures~\cite{[{See, for example, }]Rothemund2006}, and the possibility of tailoring interactions between colloidal particles to assemble diverse ordered structures and phases is also an area of active research~\cite{Sacanna2010}.

This article concentrates on self-assembly processes that occur without energy input. From a statistical mechanical viewpoint, such processes may be regarded as the relaxation of a system of interacting components towards their equilibrium state.  It is common to draw an analogy with phase changes such as crystallisation, as discussed long ago by Caspar and Klug in the context of viral capsid assembly~\cite{caspar-klug}.  From a theoretical perspective, it is natural to separate considerations of self-assembly into two parts: Firstly, the thermodynamic (or static) problem of determining what equilibrium states can be generated by varying the interactions between particles. Secondly, there are dynamic questions: how long does it take for a system to reach its ordered equilibrium state, and how can inter-particle interactions be optimised to facilitate rapid and effective assembly?

This article is concerned with these dynamic questions.  Even if the equilibrium state of a system is ordered, there are many scenarios in which formation of the relevant order occurs extremely slowly. In self-assembly, this behavior is often referred to as `kinetic trapping'; in the statistical mechanics of phase change, one more often refers to dynamical arrest or to metastable disordered states. Practically speaking, these terms indicate that self-assembly is rendered ineffective, and that equilibration of the system is very slow.

In recent years, several  studies ~\cite{Whitesides2002,Hagan2006,Jack2007,Zlotnick2007,Rapaport2008,Elrad2008,Whitelam2009,Nguyen2007,Wilber2007,Wilber2009,Klotsa2011} have observed that kinetic trapping tends to occur when interparticle bonds are very strong, and that assembly is most often effective when structures are stabilised by large numbers of relatively weak interactions. In such cases, effective self-assembly processes are characterised by transient bond formation and bond breaking events, leading to the idea that microscopic reversibility acts to facilitate effective assembly.

In this article, we use theoretical and computational methods from studies of phase change to analyse kinetic trapping in self-assembly processes.  We concentrate on two kinds of kinetic trapping, one of which is familiar from classical theories of phase change while the other is accompanied by a breakdown of the classical theories, and is accompanied by the appearance of many  long-lived disordered states.  We discuss the breakdown of these classical theories, and demonstrate that it may be identified and characterised through tests of `local equilibration' assumptions, as proposed by some of us~\cite{Jack2007}. We illustrate our analysis with a model of viral capsid assembly and a simple lattice
gas model undergoing phase separation into dense and dilute phases.

\section{Models}

\begin{figure}
\includegraphics{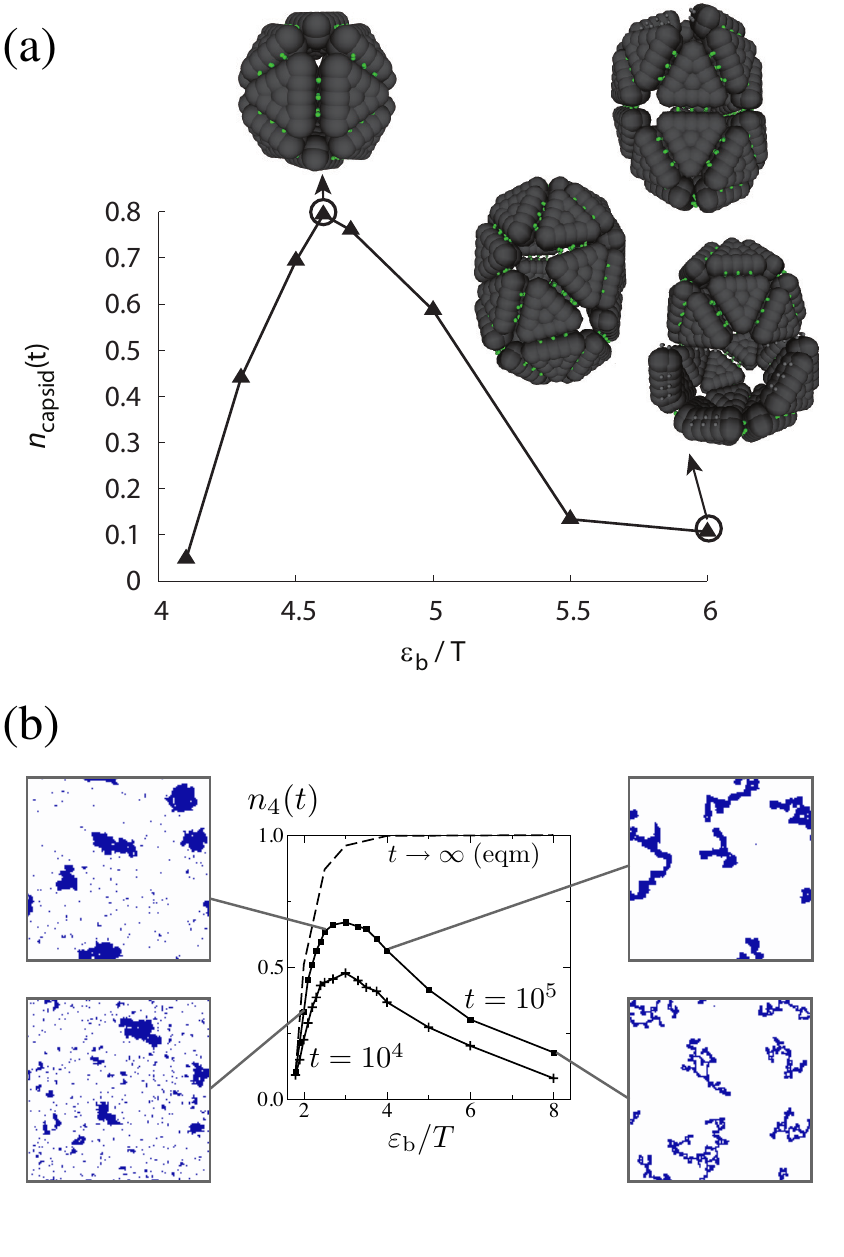}
\caption{Assembly from a disordered state.  (a) Dynamical capsid assembly yields in the $NVT$ ensemble. The fraction of subunits in well-formed capsids, $n_\mathrm{capsid}(t)$ is shown for $t=210,000t_0$ as a function of the binding interaction parameter $\eb$. Snapshots exemplify typical clusters at the circled points. Green attractor pseudoatoms are experiencing favorable interactions, while gray attractors are not. The size of the attractors indicates the length scale of their interaction. The system contains $N=500$ trimer subunits in a box of sidelength $L=74 \sigma\sub{b}$  (b) Phase change in the lattice gas at density $\rho=0.1$.  The binodal is located at $\eb/T=1.86$. The assembly yield is the fraction of particles that have four bonds $n_4(t)$.  The snapshots show representative configurations at time $t=10^5$. The dashed line shows the yield that would be obtained by equilibrating a very large system: this `thermodynamic yield' is monotonic in the bond strength $\eb$
while the yield at fixed time is non-monotonic.  The lattice size is $L=128$.
}
\label{fig:quench}
\end{figure}

\subsection{Viral capsid model}

We first describe a model for the self-assembly of empty icosahedral viral capsids. The model represents capsid proteins as rigid bodies (`subunits') with excluded volume geometries and orientation-dependent interactions.  The lowest energy structure is an icosahedral shell consisting of 20 subunits (details are given in Appendix~\ref{app:capsid} and Fig.~\ref{fig:modelCapsid} as well as in Ref.~\cite{Elrad2010}). This model was used to examine the assembly of icosahedral viruses around a polymer in Ref.~\cite{Elrad2010}, and is similar to models used by Rapaport \emph{et al.}~\cite{Rapaport1999,*Rapaport2004,Rapaport2008} and Nguyen \emph{et al.}~\cite{Nguyen2007} in simulations of empty capsid assembly. Each subunit could correspond to a `capsomer' comprising a trimer of proteins that form a $T=1$ capsid.

To simulate the dynamical process of self-assembly, we use over-damped Brownian dynamics for the capsid subunits as in Ref.~\cite{Elrad2010}, using periodic boundary conditions and a second order predictor-corrector algorithm~\cite{Branka1999,*Heyes2000}. The capsomer subunits have anisotropic translational and rotational diffusion constants calculated using Hydrosub7.C~\cite{Hydrosub}. To obtain dimensionless units, we rescale lengths by $\sigma\sub{b}$, which is the diameter of one of the spheres that comprise the excluded volume of the capsomer; times are measured in units of $t_0$, which is the Brownian time for a single such sphere.  The binding energy associated with each interaction site on a subunit is $\eb$ and we take Boltzmann's constant $k\sub{B}=1$ so that the relevant dimensionless
parameter is $\eb/T$. Further details of the model are given in Appendix~\ref{app:capsid}.

\subsection{Viral capsid assembly in the canonical ensemble.}

In Fig.~\ref{fig:quench}(a), we show results from simulations of self-assembly at constant particle number, volume, and temperature ($NVT$) for various values of the interaction energy $\eb$.  The initial conditions for the simulations have subunits with random positions and orientations.  We measure the number of perfect capsids in the system (a perfect capsid is defined as a cluster with exactly 20 subunits, each of which has its maximum number of 3 bonds). We associate the fraction of capsomer subunits in perfect capsids $n\sub{capsid}(t)$ with the \emph{yield} of the assembly process~\cite{Note1}.

As anticipated in Section~\ref{sec:intro}, the yield depends on a combination of thermodynamic and kinetic effects.  For weak bonds (small $\eb/T$), there is little thermodynamic drive to assemble and no capsids are formed. For strong bonds (large $\eb/T$), the thermodynamic drive to assembly is strong, but the system is vulnerable to kinetic trapping and forms disordered clusters of subunits instead of perfect capsids.  Optimal assembly takes place in an intermediate range $\eb/T\approx 4.5$\cite{Note2}.
In later sections, we will analyse the interplay of thermodynamic and kinetic effects in more detail.

\subsection{Lattice gas model of self-assembly}
\label{subsec:ising_model}

We also consider `self-assembly' in an Ising lattice gas containing $N$ particles on a (two-dimensional) square lattice with $V=L^2$ sites. Particles may not overlap, so the occupancy of site $i$ is $n_i\in\{0,1\}$. Particles on neighbouring sites form bonds with energy $\eb$ so that the energy of a configuration is
\begin{equation}
E = - \eb \sum_{\langle ij\rangle} n_j n_j
\label{equ:lgE}
\end{equation}
where the sum runs over (distinct) pairs of nearest neighbors.  Working with a fixed number of particles, the system is unstable to phase separation at low temperatures, forming dense (liquid) and dilute (gas) phases.  To make an analogy with self-assembly, we start with the particles in a disordered configuration, and measure the rate with which order is formed (see also~\cite{Whitelam2009}). To quantify the \emph{yield} of the assembly process, we measure the number of particles that have bonds to all four of their neighboring sites: we denote this number by $N_4$ and we write $n_4(t) = \frac1N \langle N_4(t)\rangle$ for the fraction of particles with 4 bonds.

The model evolves in time according to a Monte Carlo (MC) procedure that involves cluster moves, chosen to produce trajectories that are dynamically realistic, at least qualitatively~\cite{Whitelam2007,Whitelam2010}.
The method that we use is close to that described in Ref.~\cite{Bhatt2008}. In each MC move, we select a seed particle and use it to build a cluster, as follows. For each particle bonded to the seed particle, we conduct a Monte Carlo trial, adding it to the cluster with probability $p_\mathrm{c}=1-\ee^{-\eb/T}$. This process is then repeated recursively: for those particles that have been added, we use the same MC trial to decide whether particles bonded to them are added in turn. Taking the resulting cluster, we propose a move in a random direction. This move is rejected if this proposed move would lead to more than one particle on any site.  Otherwise, the move is accepted with a probability $p_\mathrm{a}=1/n^2$, where $n$ is the size of the cluster to be moved. An MC sweep consists of $N$ moves, and time is measured in MC sweeps. The choices of $p_\mathrm{c}$ and $p_\mathrm{a}$ ensure that the procedure obeys detailed balance
with respect to a Boltzmann distribution whose energy is given by (\ref{equ:lgE}), and also that large clusters of particles diffuse freely through the system with a diffusion constant consistent with Brownian dynamics $D(n)\propto 1/n$. This dynamical scheme represents a schematic description of particles with short-ranged attractions
moving through a solvent~\cite{Note3}.

The relevant variables in our $NVT$ simulations of this system are the dimensionless bond strength $\eb/T$ and density $\rho=N/V$. The phase behaviour as a function of these two parameters is well-known: the system is unstable to phase separation at temperatures below the binodal (that is, when $\sinh^4(\eb/2T) > 1/[1-(2\rho-1)^8]$). To obtain Fig.~\ref{fig:quench}(b), we initialise particles in random positions, and propagate the dynamics. The particles assemble into clusters: for temperatures below the binodal, these clusters will grow until their size becomes comparable
to the whole system.  However, for the times we consider, domains are much smaller than the system size, so the system is always far from equilibrium. As in the viral capsid model, the yield $n_4(t)$ is non-monotonic in the bond strength $\eb/T$: the thermodynamic driving force to assemble is small when $\eb/T$ is small, while kinetic trapping sets in for large $\eb/T$.

\begin{figure*}
\includegraphics[width=17cm]{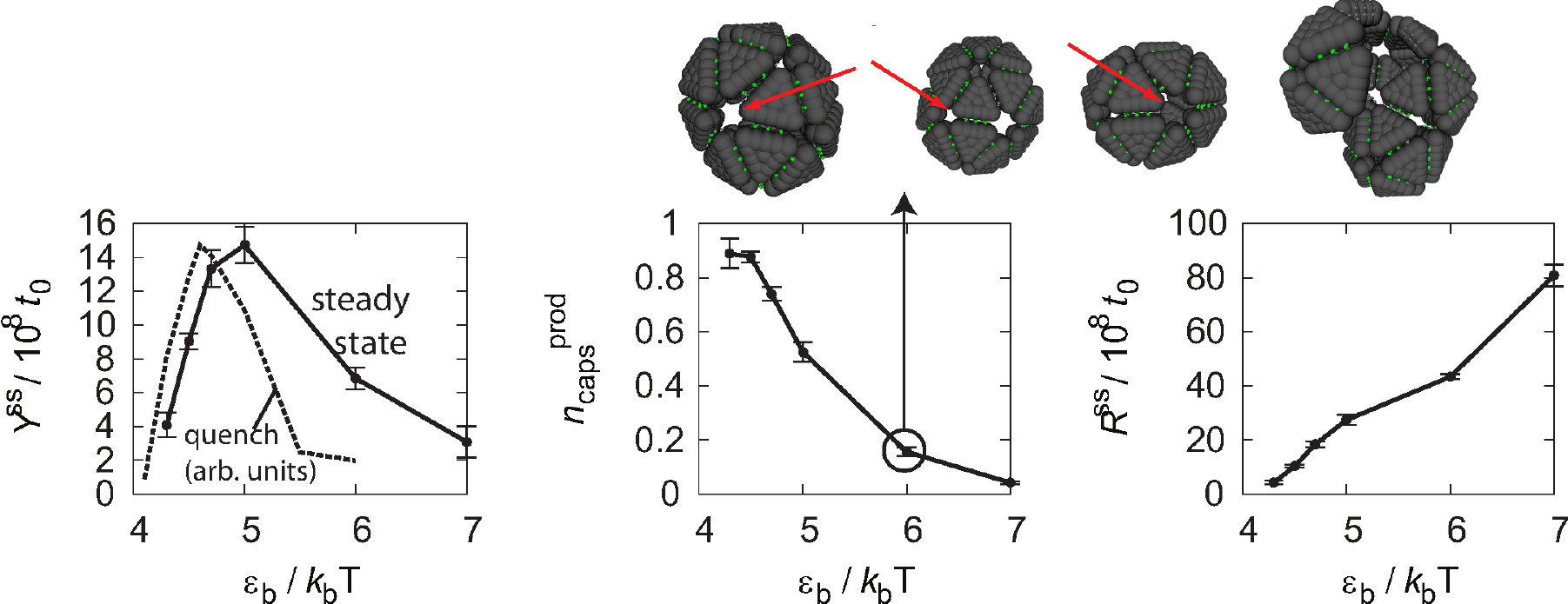}
\caption{Steady state ensemble results for the capsid model. We show the yield $Y^\mathrm{ss}$, the
product `quality' $n_\mathrm{caps}^\mathrm{prod}$ and the production rate $R^\mathrm{ss}$. Four snapshots corresponding to product clusters from the circled perimeter set, $\eb=6.0$  are shown on top of the plots. The steady state simulations had $N=1000$ trimer subunits in a box with sidelength $L=93.23 \sigma\sub{b}$. The red arrows indicate the location of hexameric defects discussed in section \ref{subsec:morph}.}
\label{fig:caps_mallet}
\end{figure*}

\section{Steady state ensemble -- rate and quality of assembly}
\label{sec:mallet}

\subsection{Steady state ensemble}
\label{subsec:mallet_general}

We now discuss a self-assembling steady state, using ideas that have been exploited in studies of nucleation and phase transformation~\cite{Maibaum2008}. Given a self-assembling model system such as the capsid or lattice gas model, we simulate the dynamics of the system in the usual way, except that we periodically remove large clusters of particles (subunits) from the system.  We refer to clusters removed in this way as the \emph{products} of the self-assembly process. The morphologies of the product clusters are stored for later analysis, and we then re-introduce free particles into the system at random positions so that the total number of particles in the system remains constant. To make connection with experiment, we imagine a continuous assembly process, where free particles (subunits) are fed into the system and large assembled products are removed, perhaps by exploiting their tendency to sediment. On starting the system in an initially random configuration, it settles down into a time-translationally invariant steady state in which product clusters are continuously assembling. This feature of the steady state ensemble allows time-invariant averages to be taken during productive assembly, in contrast to the NVT quenches.  The criteria for identifying large clusters depend on the model being simulated and are described below.

We use the notation $\langle \cdot \rangle_\mathrm{ss}$ for averages within the steady state.  For example, the average lattice gas energy $\langle E(t) \rangle_\mathrm{ss}$ is obtained by averaging the energy defined in~(\ref{equ:lgE}) at a time $t$ in the steady state regime. We also take averages over the product clusters that are formed in the steady state.  To be precise, after a simulation has been in the steady state for a time $\tobs$, let the number of product clusters formed in that time be $\cal M$.  Averaging over many such runs, we obtain the rate of product formation, per unit time and unit volume
\begin{equation}
R^\mathrm{ss} \equiv \frac{1}{V\tobs} \langle {\cal M} \rangle.
\end{equation}
Labelling the product clusters from a given run by an index $\mu=1,2,\dots,\mathcal{M}$, we may then calculate averages over these product clusters, which we denote by $\langle \cdot \rangle_\mathrm{prod}$.  For example, in the lattice gas, if $N(\mu) $ is the number of particles in cluster $\mu$ then $\langle N(\mu) \rangle_\mathrm{prod} $ is obtained by averaging this number over all product clusters.

To make a connection between the steady state ensemble and the $NVT$ simulations of Fig.~\ref{fig:quench}, it is useful to define the steady state yield
\begin{equation}
Y^\mathrm{ss} \equiv Q^\mathrm{prod} \times  R^\mathrm{ss}
\end{equation}
where $Q^\mathrm{prod}$ is a measure of the `assembly quality' of the products.  For example, in the viral capsid
model $Q^\mathrm{prod}$ is the fraction of product clusters that are perfect capsids, so that the steady state yield $Y^\mathrm{ss}$ is the production rate of perfect capsids.

\subsection{Viral capsid model in the steady state ensemble}

Results from the steady state ensemble of the viral capsid model are shown in Fig.~\ref{fig:caps_mallet}.
In this model, we define particles to be \emph{connected} if they enjoy a bond with interaction strength $U< -5 \kt$.
A cluster of connected particles is identified as a \emph{product} if (i) the cluster has
either more than 22 particles or it is a perfect capsid,
and (ii) the cluster has remained bonded with at least 17 of the same subunits for a time $t \ge 13.3 t_0$. We have in mind that the products of the assembly process be stable long-lived clusters and this second condition limits erroneous identification of weakly-bonded short-lived clusters as products.

In Fig.~\ref{fig:caps_mallet}, we show the production rate $R^\mathrm{ss}$, the yield $Y^\mathrm{ss}$, and the product quality $Q^\mathrm{prod}$, which is equal to the fraction of product clusters that are perfect capsids $n\sub{caps}^\mathrm{prod}$. As for the $NVT$ simulations of Fig.~\ref{fig:quench}, we observe that the yield is nonmonotonic with respect to binding energy, with an optimum at $\eb/T\approx 5$. The origin of this optimum is a competition between a rate $R^\mathrm{ss}$ that increases on increasing $\eb$, and a quality factor $Q(\mu)\rangle\sub{prod}$ that decreases.  (The total production rate increases with $\eb$ over the whole range considered, although it eventually decreases at much higher $\eb$, for reasons discussed in section~\ref{sec:rate} below.)

In terms of kinetic trapping, we find that for large $\eb/T$ product clusters are being formed quickly, but
these clusters are of low quality.  In later sections we contrast this scenario with the `stalling' or `starvation'
scenarios discussed by Zlotnick~\cite{Endres2002} in the context of viral capsid assembly.  In that case, kinetic
trapping appears as a rate $R^\mathrm{ss}$ that decreases sharply as $\eb/T$ is large (see also Sec.~\ref{sec:rate} below). Fig.~\ref{fig:caps_mallet} shows that this is not the case for the steady state ensemble with the parameters we simulate here.

\subsection{Lattice gas model in the steady state ensemble}

\begin{figure*}
\includegraphics[width=15cm]{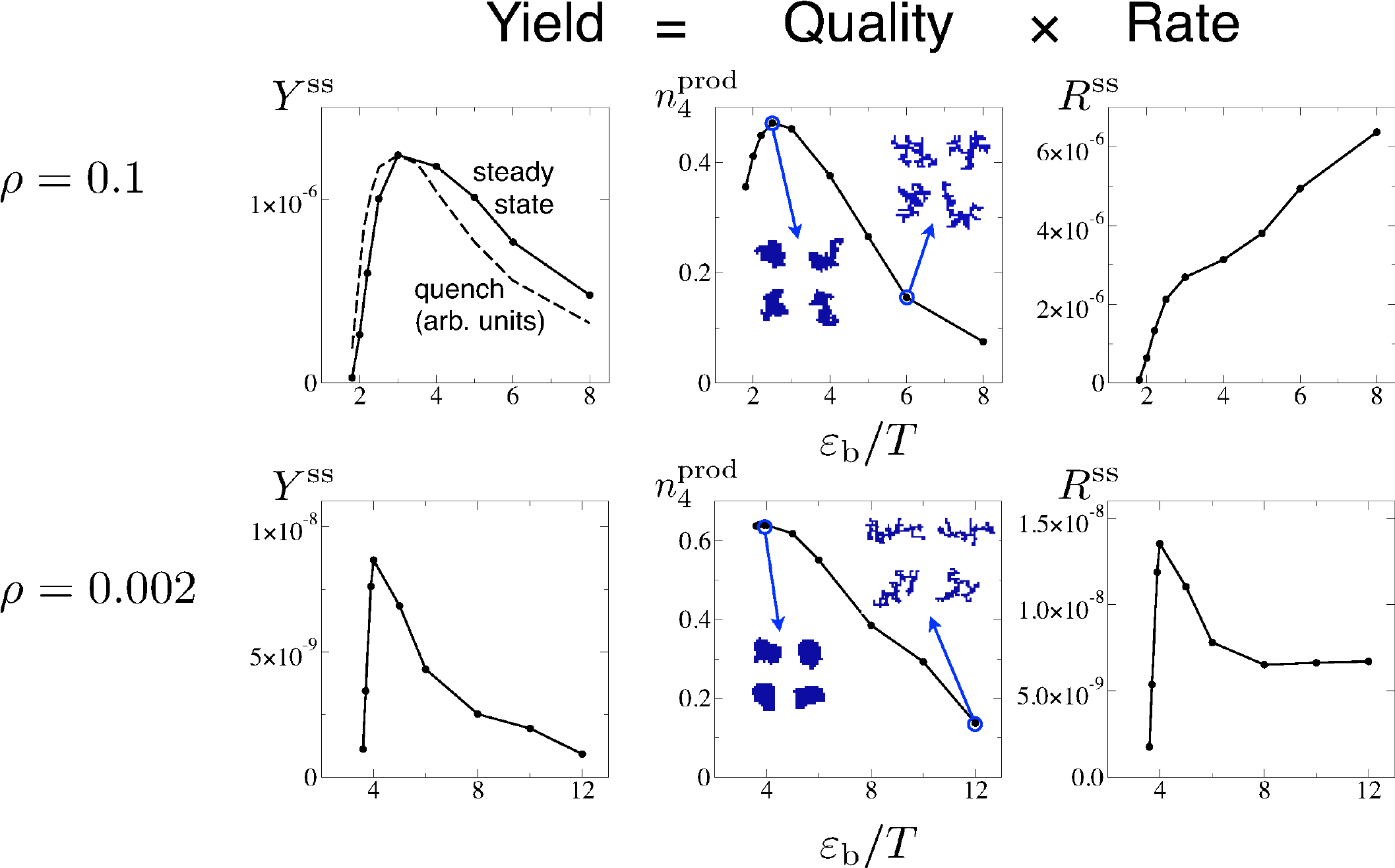}
\caption{
Steady state ensemble in the lattice gas with $\nm=100$.
At two different densities, we show the yield $Y^\mathrm{ss}$, the product `quality' $n_4^\mathrm{prod}$ and the `production rate' $R^\mathrm{ss}$.  At $\rho=0.1$, we compare with the yield $\langle n_4(t)\rangle$ at a time $t=10^5$ after a `quench' from a disordered state (data from Fig.~\ref{fig:quench}, rescaled for comparison).  In the central panels, we show example `product' clusters, to indicate their morphologies.}
\label{fig:is_mallet}
\end{figure*}

In the lattice gas model, clusters are identified as products if their size is larger than a maximal cluster size $\nm$.  We choose $\nm=100$ although our results depend only weakly on $\nm$. For the product quality $Q^\mathrm{prod}$, we take the fraction of product particles that have 4 bonds, calculated as
\begin{equation}
Q^\mathrm{prod} = n_4^\mathrm{prod} =
\frac{ \langle N_4(\mu) \rangle_\mathrm{prod} } { \langle N(\mu) \rangle_\mathrm{prod} }
\end{equation}
where $N_4(\mu)$ is the number of 4-bonded particles in product cluster $\mu$.

Results from the steady state ensemble are shown in Fig.~\ref{fig:is_mallet} for two different densities $\rho$. The non-monotonic behaviour of the yield $Y^\mathrm{ss}$ mirrors the behaviour of the yield $n_4(t)$ found in Fig.~\ref{fig:quench}~\cite{Note4}.
At $\rho=0.1$ (top panels of Fig.~\ref{fig:is_mallet}) the results are similar to those shown for the capsid system in Fig.~\ref{fig:caps_mallet}:
on increasing $\eb/T$, the non-monotonic yield arises from a competition between an increasing production rate $R^\mathrm{ss}$ and a decreasing quality $Q^\mathrm{prod}$. However, in simulations at a lower density $\rho=0.002$, the rate $R^\mathrm{ss}$ is itself non-monotonic. The scenario that occurs at low densities is consistent with a `stalling' effect~\cite{Endres2002}, where the system is depleted of free particles, leading to slow cluster growth.  But it is the relatively high density ($\rho=0.1$) scenario in the lattice gas that mimics the data for the viral capsid model shown in Fig.~\ref{fig:caps_mallet}.  As in that case, kinetic trapping occurs not just because of depletion of free particles, but rather from disordered large clusters or aggregates, examples of which are shown in Fig.~\ref{fig:is_mallet}.

\section{Measures of cluster equilibration}
\label{sec:clust_eq}

We now discuss the relation between cluster quality $Q^\mathrm{prod}$ and a condition that we call `cluster equilibration'.  Our idea is that one type of kinetic trapping arises from disordered aggregates such as those discussed
above, and that the importance of these aggregates may be measured through deviations from cluster equilibration\cite{Note7}.

For a general definition of cluster equilibration,we characterise clusters of particles by their size $n$ and by a second index $\alpha,\beta,\gamma,\dots$ that indicates their morphology.  If $\mathcal{N}_{n,\alpha}$ is the number of clusters with size $n$ and morphology $\alpha$ then our cluster equilibration condition is
\begin{equation}
\frac{\langle \mathcal{N}_{n,\alpha}\rangle}{\langle \mathcal{N}_{n,\gamma}\rangle} = \ee^{-(E_{n,\alpha} - E_{n,\gamma})/T}
\label{equ:clust_eq}
\end{equation}
where $E_{n,\alpha}$ is the energy of a cluster of $n$ particles and morphology $\alpha$, and the averages might be taken at a fixed time during assembly in the  $NVT$ ensemble, or in the steady state ensemble. In words, (\ref{equ:clust_eq}) states that: `for clusters of size $n$, the probabilities of different morphologies are Boltzmann-distributed'.  It seems natural to interpret this as a `cluster equilibration' condition. (If the clusters have different excluded volumes one might take this into account by replacing the energy with a suitable enthalpy, and any internal entropy of the cluster can also be incorporated through a cluster free energy.) In theoretical treatments of self-assembly based on rate equations or field-theoretic arguments, it is natural to assume that (\ref{equ:clust_eq}) holds (see Sec.~\ref{sec:rate} below).  We now show that deviations from (\ref{equ:clust_eq}) are signficant throughout the regimes where kinetic trapping is important, indicating that such deviations must be taken into account in theories of self-assembly.

\subsection{Cluster equilibration in the lattice gas model}

\begin{figure}
\begin{center}
\includegraphics{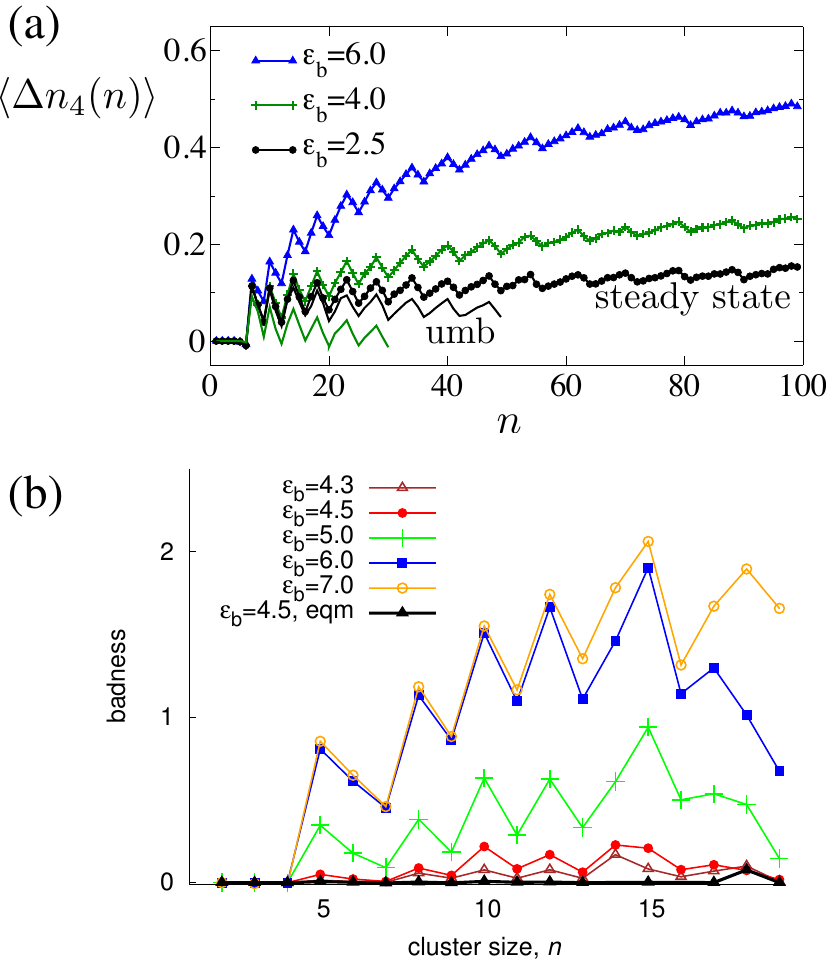}
\end{center}
\caption{(a)~Measurement of cluster equilibration in the lattice gas model at $\rho=0.1$, by comparison
of steady state and umbrella-sampled ensembles.  The data points
show $\langle \Delta n_4(n)\rangle\sub{ss}$ while the solid lines show $\langle \Delta n_4(n)\rangle\sub{umb}$.
For $\eb/T=2.5$, the system is far from global equilibrium, but deviations from the cluster
equilibration condition are small (compare the black symbols with the solid black lines).  As $\eb/T$
increases, deviations from cluster equilibration increase.
(b)~Similar data for the viral capsid model.  The deviation in number of bonds between clusters and their ground states is shown as a function of the cluster size $n$ for indicated values of $\eb$. All data points correspond to results from the steady state ensemble except for the curve with black $\blacktriangle$ symbols, which were obtained from umbrella sampling.}
\label{fig:is_quality}
\label{fig:capsidBondsOP}
\end{figure}

In the self-assembling steady state of the lattice gas model, we count the number of four-bonded particles
within each cluster. We average this quantity over clusters of a fixed size $n$, and we denote this average by
$\langle N_4(n) \rangle_\mathrm{ss}$.  We emphasise that these
are averages over clusters in the self-assembling steady state, and not over product clusters.
It is convenient to compare  $\langle N_4(n) \rangle_\mathrm{ss}$ with $N_4^\mathrm{gs}(n)$, which is the number
of four-bonded particles in a cluster of size $n$ that minimises the cluster energy.
We then define
\begin{align}
\langle \Delta n_4(n)\rangle_\mathrm{ss} &= \frac{1}{n} \langle N_4(n) - N_4^\mathrm{gs}(n) \rangle_\mathrm{ss}
\label{equ:lgbad}
\end{align}
to measure the deviation of the cluster `quality' from its ground state value, normalised by the cluster size $n$.

To test the extent of cluster equilibration, we have performed umbrella sampling, in which we choose a maximal
cluster size $n_\mathrm{umb}$ and reject all MC moves that form clusters of size bigger than $n_\mathrm{umb}$.
On propagating the dynamics, the system relaxes to a state that satisfies this constraint but is
otherwise equilibrated, so that we expect (\ref{equ:clust_eq}) to hold~\cite{Note5}.

In the umbrella-sampled ensemble, we again measure the number of particles with four bonds and
average over clusters of size $n$.  The analogue of (\ref{equ:lgbad}) within this ensemble is
$
\langle \Delta n_4(n)\rangle_\mathrm{umb} = \frac{1}{n} \langle N_4(n) - N_4^\mathrm{gs}(n) \rangle_\mathrm{umb}
$.
Comparison of $\langle \Delta n_4(n)\rangle$ between ensembles allows a test of the cluster equilibration condition: if  (\ref{equ:clust_eq}) holds exactly in the self-assembling steady state then $\langle \Delta n_4(n)\rangle_\mathrm{ss} = \langle \Delta n_4(n)\rangle_\mathrm{umb} $. In Fig.~\ref{fig:is_quality}, it can be seen that cluster equilibration holds quite accurately at $\eb/T=2.5$ which is close to the maximum of the yield (recall Figs.~\ref{fig:quench}(b) and~\ref{fig:is_mallet}).  However, as $\eb/T$ increases and assembly quality is reduced, a strong departure from cluster equilibration is apparent: we find that $\langle\Delta n_4(n)\rangle_\mathrm{ss}$increases while $\langle\Delta n_4(n)\rangle_\mathrm{umb}$ decreases.  The key point is that the crossover in $Q^\mathrm{prod}$ in Fig.~\ref{fig:is_mallet} and the deviations from cluster equilibration occur at similar values of the bond strength. Our conclusion is that effective self-assembly requires transient bond-breaking processes in order to avoid kinetic trapping, and further that these bond-breaking processes need to be frequent enough that the system is close to the cluster equilibration condition~(\ref{equ:clust_eq}).

Finally, we note that $\langle \Delta n_4(n) \rangle$ tends to increase with $n$ in a `sawtooth' fashion.  The effect is primarily due to  the quantity $N_4^\mathrm{gs}(n) $ that appears in the definition of $\langle \Delta n_4(n) \rangle$.  As $n$ increases, $N_4^\mathrm{gs}(n)$ changes in discrete steps of various sizes, depending on the precise nature of the cluster ground state. However, there are often a range of cluster morphologies with energies close to the ground state energy, all of which occur with significant probability in both umbrella-sampled and steady state ensembles.  The effect of these clusters is that $\langle N_4(n)\rangle$ depends more smoothly on $n$ than $N_4^\mathrm{gs}(n)$, resulting in a sawtooth structure in $\langle \Delta n_4(n)\rangle$.  For our purposes, the relevant comparison is between umbrella-sampled and steady-state data, which both exhibit similar $n$-dependence in this case.

\subsection{Cluster equilibration in the viral capsid model}

To test cluster equilibration in the viral capsid model, we concentrate on the average number of bonds in clusters of size $n$, denoted by $\langle B(n) \rangle_\mathrm{ss}$. We compare this average with the number of bonds $B^\mathrm{gs}(n)$ in a cluster of size $n$ with minimal energy. In this case the absolute deviation from the ground state cluster is of particular relevance, since rate equation descriptions of capsid assembly often consider only the ground state morphology for each intermediate size. Therefore we define
\begin{equation}
\bad_\mathrm{ss} = \langle B(n) - B^\mathrm{gs}(n) \rangle_\mathrm{ss}.
\end{equation}
Note that this deviation is not normalized by the cluster size $n$. As for the lattice gas, we perform umbrella sampling that prohibits the formation of clusters larger than $n\sub{umb}$. (Specifically, we use a hybrid Brownian dynamics/Monte Carlo approach where we use a short sequence of unbiased Brownian dynamics steps as a trial move, which is rejected if the size of the largest cluster is greater than $n\sub{umb}$.)

Results for $\bad$ are shown in Fig.~\ref{fig:capsidBondsOP}. For the umbrella-sampled data, we find that $\bad\sub{umb}\approx0$ for $\eb=4.5$: this quantity is similarly small for $\eb>4.5$. (The apparent deviation from $\bad_\mathrm{umb}\approx0$ at $n=18$ in Fig.~\ref{fig:capsidBondsOP} is likely a result of imperfect equilibration in the umbrella sampled simulations). As in the lattice gas data, the steady state measurements show that deviations from cluster equilibration are small near optimal assembly, and grow as kinetic trapping sets in and $Q^\mathrm{prod}$ decreases.

As in the lattice gas, a sawtooth structure is visible in $\bad\sub{ss}$.  Here, increasing the cluster size $n$ leads to a change of either one or two bonds in $B^\mathrm{gs}(n)$.  As $\bad\sub{ss}$ deviates from $B^\mathrm{gs}(n)$, there are several relevant cluster morphologies in the steady state ensemble which average out the step changes that occur in $B^\mathrm{gs}(n)$: typically, the change in $\langle B(n)\rangle\sub{ss}$ on increasing $n$ would be somewhere between 1 and 2 bonds. The combination of discrete changes in $B^\mathrm{gs}(n)$ and smoother changes in $\langle B(n) \rangle \sub{ss}$ results in the apparent sawtooth pattern.

\begin{figure}
\includegraphics{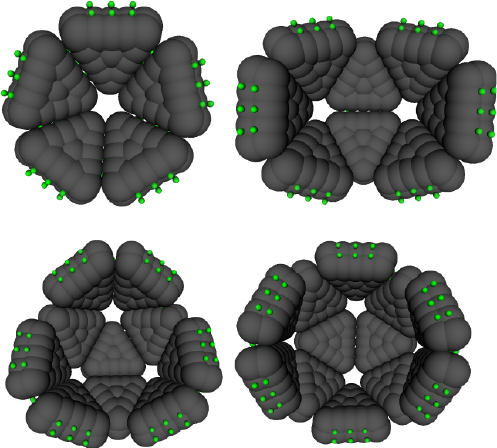}
\caption{Clusters of sizes (top left to bottom-right) 5, 8, 10, and 12 in which every capsomer has at least two bonds. For numbers of subunits in between the sizes shown, there are no structures in which every capsomer has at least two unstrained bonds. This pattern gives rise to the  sawtooth form for $\bad$ in Fig.~\ref{fig:capsidBondsOP}.}
\label{fig:closedpoly}
\end{figure}

Results in the umbrella sampled ensemble are analysed in more detail in Appendix~\ref{app:freeEnergies}.  We find that the free energy of a cluster of size $n$ can be obtained by analysing the total number of bonds formed together with the entropy associated with different ground state morphologies.  For the purposes of this section, the sawtooth structure in Fig.~\ref{fig:capsidBondsOP}(b) can be attributed to the fact that some cluster sizes ($n=5,8,10,\dots$) have ground states in which every capsomer has at least two bonds (see Fig.~\ref{fig:closedpoly}). For these structures $B^\mathrm{gs}(n)$ is large, but the number of such morphologies is rather small (see in particular Fig.~\ref{fig:freeEnergy}(b)).  As cluster equilibration breaks down, the effect on $\bad$ is most pronounced for these cluster sizes, since there are many available morphologies with fewer bonds than the ground state, and these morphologies tend to form most quickly as clusters grow.  For other cluster sizes, deviation from cluster equilibrium are less pronounced, since there are diverse ground state morphologies, all of which are kinetically accessible.

\section{Kinetic equations in self-assembly}
\label{sec:rate}

In the capsid and lattice gas models, clusters of particles grow as assembly takes place.  A natural approach is therefore to describe this process in terms of kinetic rate equations for cluster concentrations.  In phase change processes, this idea goes back to Becker and D\"oring~\cite{Becker1935}, and a derivation of this approach from the microscopic dynamics of the lattice gas (or Ising) model was considered by Binder and Stauffer~\cite{Binder1976}.  Similar ideas have been developed by Zlotnick and co-workers in order to describe viral capsid assembly~\cite{Zlotnick2005}.

In this section, we show that non-monotonic steady-state yields $Y^\mathrm{ss}$ can be predicted by such equations, but we emphasise that these equations fail to capture the decreasing quality $Q^\mathrm{prod}$ that occurs in both capsid and lattice gas models.  We argue that this failure of kinetic rate equations is linked with the breakdown of the cluster equilibration condition (\ref{equ:clust_eq}).

\subsection{Equations for cluster growth and self-assembly}
\label{sec:rate_ass}

The central idea behind kinetic rate equations is to organise configurations of the system according to the sizes of the clusters  that are present in the system.  Let $\mathcal{N}_n(t)$ be the number of clusters of size $n$, at some time $t$, so that $\rho_n(t)=\mathcal{N}_n(t)/V$ is the concentration of such clusters. For large systems where the various clusters are well-mixed and interact through binary collisions, one often writes
\begin{align}
\frac{\partial}{\partial t}\rho_n(t) =& \sum_{n'} [ W^+_{n-n',n'} \rho_{n-n'}(t) - W^+_{n,n'} \rho_n(t) ] \rho_{n'}(t)
 \nonumber \\ & +  \sum_{n'} [W^-_{n+n',n'} \rho_{n+n'}(t) - W^-_{n,n'}] \rho_{n}(t)
\label{equ:clust_master}
\end{align}
where the coeffecients $W^+$ and $W^-$ are rate constants for binary cluster fusion and cluster fission events respectively. We use a notation where the sums over $n'$ are unrestricted, but the coefficients $W^\pm_{n,n'}$ are zero for $n'\geq n$. For a simple description, we may take $W^+_{n,n'}$ and $W^-_{n,n'}$ to be finite only when $n'=1$, recovering the classical Becker-D\"oring equation.

The restriction to binary collisions may be relaxed straightforwardly (see, for eample~\cite{Maibaum2008}) and cases when the clusters are not well-mixed can be treated by field-theoretic approachs~\cite{Bray1994}.  However, an additional assumption on writing (\ref{equ:clust_master}) is that \emph{all clusters of size $n$ behave statistically identically, regardless of their shape.}  This assumption is tied in with our condition of cluster equilibration above, as discussed in Sec.~\ref{subsec:morph} below.

\subsection{Non-monotonic production rate $R$ in kinetic equations}

The steady state ensemble has a natural realisation in terms of these kinetic rate equations. To keep a compact notation, we write $M=\nm-1$ as the size of the largest clusters that are not removed as products. We consider clusters of sizes $n=1\dots M$, and we restrict ourselves for convenience to monomer binding and unbinding. Then, for $1<n<M$ we have
\begin{equation}
\frac{\partial}{\partial t} \rho_n(t) = D \rho_1(t) [ \rho_{n-1}(t) - \rho_n(t) ] + \lambda_{n+1} \rho_{n+1}(t) - \lambda_n \rho_n(t)
\label{equ:dtrho}
\end{equation}
For simplicity, we have replaced the $n$-dependent rate constants by a single `diffusion-limited' rate
$D$\cite{Note6},
and $\lambda_n$ is the rate for unbinding of a monomer from a cluster of size $n$.
If the system is allowed to equilibrate, we have that $\rho_n^\mathrm{eq} = \rho_1^\mathrm{eq} \prod_{n=2}^N (D\rho_1^\mathrm{eq}/\lambda_n)$.
In comparing with lattice gas or capsid models, we expect monomer binding and unbinding rates to be related by detailed balance, as
\begin{equation} D = \lambda_m v \ee^{-\eb/T}, \label{equ:D-lam} \end{equation}
where $v$ is an entropic factor associated with bonding, with dimensions of volume (specifically, the contribution of monomer attractions to the 2nd virial coefficient of the system is $v\ee^{-\eb/T}$).

In the assembling steady state, the equations of motion for $\rho_M(t)$ and $\rho_1(t)$ are modified to include the removal of product clusters: details are given in Appendix~\ref{app:mall}.  The total number of particles (subunits) in the system is a constant $\rho_\mathrm{T}=\sum_n n\rho_n$. The production rate may also be identified as $R(t) = D\rho_1(t)\rho_M(t)$.

The simplest case is irreversible binding, where bonds never break, so $\lambda_m=0$ for all $m$. As shown in Appendix~\ref{app:mall} the exact result is
\begin{equation}
R^\infty=D\rho_1\rho_M=\frac{4D\rho_T^2}{M^2(M+1)^2}.
\label{equ:Rinf}
\end{equation}
(Within the steady state, we drop all time arguments on $\rho_n$ and $R$.) The signature of kinetic trapping will be that introducing some non-zero unbinding rates $\lambda_n$ will lead to an increase in $R$ (holding $\rho_T$ constant).  That is, increasing the rate of monomer unbinding increases the production rate $R$. This is the `stalling' (starvation) effect of Zlotnick and co-workers~\cite{Endres2002}.

\begin{figure}
\includegraphics[width=8cm]{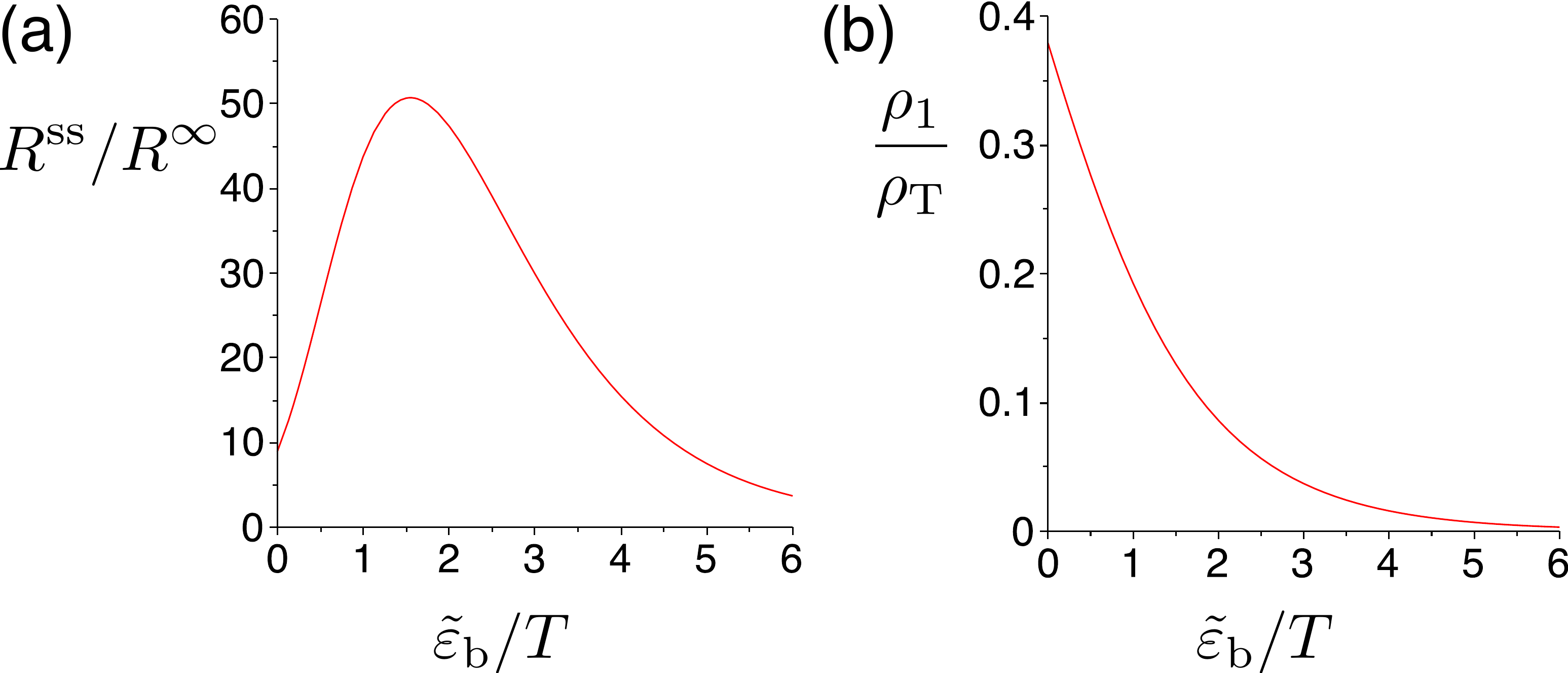}
\caption{(a)~Rate $R^\mathrm{ss}$ vs $\ebt/T$ for kinetic rate equations, showing
non-monotonic behaviour due to `kinetic trapping' in states with many intermediates and few monomers.
We take $M=50$, $m^*=10$ and the rate is normalised by its value as $\ebt\to\infty$.
(b)~The fraction of particles in the
assembling steady state that are free monomers,
 further emphasising that the small rate for large $\ebt$ arises from states with
a small number of monomers, and hence a small rate of bond formation.}
\label{fig:rate_analytic}
\end{figure}

To observe this effect in the steady state,  an essential model ingredient is that the unbinding rates $\lambda_m$ depend on the cluster size $m$.  For simplicity and to maintain contact with Refs.~\cite{Endres2002,Hagan2010}, we suppose that there is a `critical cluster size' $m^*$ above which unbinding is slow $\lambda_m\approx0$ while for small clusters we take a finite value $\lambda_m=\lambda$.  (The critical cluster size should be interpreted in the spirit of classical nucleation theory~\cite{Binder1976}.)

The production rate $R^\mathrm{ss}$ depends on $m^*$, $\nm$ and a dimensionless parameter $\lambda/D\rho_\mathrm{T}$. This last parameter determines the rate of bond-breaking for clusters with $m<m^*$: it is convenient to express this as an `effective bond strength'
\begin{equation}
\ebt/T = \log(D\rho_\mathrm{T}/\lambda_m)
\end{equation}
Comparison with (\ref{equ:D-lam}) shows that $\ebt = \eb - T\log(v\rho_\mathrm{T})$ and is thus a grand free energy with $\rho_\mathrm{T}$ the concentration of a subunit bath. That is, the relevant binding free energy depends on the total subunit density as well as the bonding parameters $\eb$ and $v$.  The key point is that within the rate equation treatment, the full dependence of the system on $\lambda$ and $\rho_\mathrm{T}$ can be obtained through the single parameter $\ebt/T$.  (We emphasise however that we have assumed that unbinding from large clusters is very slow: the rate $\lambda$ in this analysis is the rate of unbinding from \emph{small} clusters.)

The central result of this analysis is shown in Fig.~\ref{fig:rate_analytic}: the production rate $R$ shows a non-monotonic dependence on $\ebt$.  Since we are working at fixed $\rho_\mathrm{T}$, this corresponds to a non-monotonic dependence on $\eb$ in the capsid and lattice gas models. Hence, these results qualitatively mirror the behaviour shown in Fig.~\ref{fig:is_mallet} as well as the stalling (or starvation) effects discussed by Zlotnick~\cite{Endres2002}. We have verified that the non-monotonicity survives on introducing small finite rates for unbinding from large clusters (finite $\lambda_m$ for $m>m^*$), although a monotonic response is recovered if the unbinding rate is completely independent of $m$.

Physically, the interpretation of this ``starvation'' regime is that a small unbinding rate $\lambda$ acts to reduce the concentration of monomers $\rho_1$ since free subunits quickly join growing clusters.  The production rate is $R=D\rho_1\rho_M$ so a small concentration $\rho_1$ reduces this rate strongly.  As the unbinding rate $\lambda$ increases, $\rho_1$ increases quickly, while the effect on the concentration of large clusters $\rho_M$ is much weaker.  Thus $R$ increases as $\lambda$ is increased, demonstrating that kinetic trapping occurs.

We note that while we have analysed these kinetic equations in the steady state ensemble, similar non-monotonic production rates are observed on starting with disordered states and waiting for clusters to form~\cite{Zlotnick1999,Endres2002,Hagan2006,Hagan2010}.

\subsection{Cluster equilibration}
\label{subsec:morph}

The previous results demonstrate that Eqs.~(\ref{equ:clust_master}) reproduce one feature of the lattice gas and capsid models, the nonmonotonic production rate. However, it is clear from (\ref{equ:clust_master}) that this rate equation approach treats all clusters of a given size on the same footing. As discussed above, these approximations are justified if all clusters of a given size behave statistically identically.   Classically~\cite{Bray1994}, the argument supporting this assumption is that large clusters are rare, and that transitions between different morphologies are rapid compared to collisions between clusters.  If this separation of time scales holds, one may consider each cluster as a separate subsystem, which relaxes quickly to a `quasiequilibrium' state: the cluster equilibration condition (\ref{equ:clust_eq}) then holds exactly.  In practice, the condition of cluster equilibration is much weaker than the assumption of a clear separation of time scales between cluster rearrangement and cluster growth -- but the results of Sec.~\ref{sec:mallet} and \ref{sec:clust_eq} show that it is the cluster equilibration condition that breaks down as assembly quality falls.

Therefore, when modelling assembly with rate equations of this form, there is an implicit assumption that cluster equilibration holds, and hence that the assembly quality $Q^\text{prod}$ is independent of temperature.  From Figs.~\ref{fig:caps_mallet} and \ref{fig:is_mallet}, this assumption is not valid once kinetic trapping sets in. Thus, while kinetic rate equations can can reproduce a non-monotonic dependence of production rate on bond strength, our results from the steady state ensemble show clearly that these equations miss an important part of the story: the decrease of production quality as bonds get strong.

We note that there are two mechanisms by which cluster equilibration can be violated. In the first, subunits form strong interactions with a sub-optimal number of partners. In other words, each subunit-subunit interaction approximately corresponds to a minimum in the interaction potential, but subunits do not add on to a growing cluster in locations that offer the most interaction partners.
In the second mode of violation, subunits form strained bonds which deviate from the ground state of the interaction potential. For example, assembling capsids frequently form hexameric defects, as illustrated in   Fig.~\ref{fig:caps_mallet}. The first form of cluster equilibration violation can be incorporated into the rate equation approach, at a cost of significantly increased computational complexity, if the space of all possible cluster configurations can be predefined, and then relevant cluster configurations can be enumerated ahead of time \cite{Moisant2010} or sampled stochastically \cite{Sweeney2008}. However, these approaches have not been used to address the possibility of defective bonds, for which it is not possible to predefine the space of possible cluster configurations.

\section{Discussion and outlook}

The usefulness of weak interparticle bonds for self-assembly has been commented on by several authors~\cite{Hagan2006,Jack2007,Whitesides2002,Ceres2002,Zlotnick2007,Rapaport2008,Whitelam2009,Elrad2008,Elrad2010,Hagan2010}. Thermal fluctuations allow these bonds to be broken: we have shown that this effect can enhance assembly by increasing the concentration of free particles and hence the rate of cluster formation. These results are consistent with studies by Zlotnick and co-workers.  However, our simulations also identify a second mechanism by which weak bonds enhance the assembly of clusters with a given morphology. Namely, bond-breaking processes act to promote cluster equilibration, in the sense of (\ref{equ:clust_eq}). The qualitative importance of cluster equilibration  was first raised  by Whitesides and Boncheva \cite{Whitesides2002}; we have attempted to quantify this idea through Eq. (\ref{equ:clust_eq}).

The importance of kinetic trapping to biological assembly, and the constraints it places on interactions between the constituents, has been  vividly demonstrated through experiment (e.g. Refs.\cite{Ceres2002,Zlotnick2007,Katen2010}) and modeling \cite{Hagan2006,Jack2007,Endres2002,Rapaport2008,Elrad2008,Hicks2006,Nguyen2007}. If we are to anticipate the design of functionalised particles that self-assemble into ordered structures, the possibility of kinetic trapping must surely be taken into account for these systems as well.  In particular, methods for predicting the ``optimal weakness'' of interparticle bonds could streamline the design process.  In~\cite{Jack2007}, we proposed that the degree of cluster equilibration (or local equilibration) might be measured using fluctuation theorems that couple to the reversibility of bond-formation.

Developments in this direction will be discussed in future publications: here we note that the cluster equilibration condition (\ref{equ:clust_eq})is weaker than the `local equilibrium' conditions  discussed in~\cite{Jack2007}. For example, (\ref{equ:clust_eq}) may hold even in the absence of good-mixing conditions, which lead to a deviation from local equilibrium in the sense of~\cite{Jack2007}.  This distinction emphasises the point that, while some degrees of freedom in out-of-equilibrium systems may be locally equilibrated in this sense, other degrees of freedom may be far from equilibrium. For example, the recent results of Russo and Sciortino~\cite{Russo2010} seem to indicate that density fluctuations are much closer to a local equilibrium distribution than energy fluctuations.  We conjecture that the near-local equilibration of density is linked with a weak violation of the good-mixing assumption, while the energy fluctuations reflect  a stronger violation of cluster equilibration, in this out-of-equilibrium system.

More generally, we conclude that our results are entirely consistent with the general idea~\cite{Whitesides2002} that effective self-assembly occurs through the reversible formation of numerous weak bonds.  We believe that statistical mechanical methods can be used to test this idea quantitatively, with a view to exploiting it in the design and control of self-assembly process.  In particular, the breakdown of cluster equilibration when bonds are strong is a kinetic effect that is not taken into account in classical theories of self-assembly and phase change.  We believe that the development of quantitative methods for characterising this effect is a key challenge for theoretical studies of self-assembly, and we look forward to further progress in this area.

\begin{acknowledgments}
We thank Steve Whitelam, Phill Geissler, and David Chandler for many discussions on the importance of reversibility
in self-assembly,
and RLJ thanks Stephen Williams for helpful discussions of local equilibration and quasi-equilibrium.
This work was supported by Award Number R01AI080791 from the National Institute Of Allergy And Infectious Diseases (to MFH and OME) and by the EPSRC through grants EP/G038074/1 and EP/I003797/1 (to RLJ).  MFH also acknowledges support by National Science Foundation through the Brandeis Materials Research Science and Engineering Center (MRSEC). Computational resources were provided by the National Science Foundation through TeraGrid computing resources (specifically the Purdue Condor pool) and the Brandeis HPCC.
\end{acknowledgments}

\begin{appendix}
\section{Description of the capsid model}
\label{app:capsid}
The model subunits are comprised of a set of overlapping spherical `excluders' that enforce excluded volume and spherical `attractors' with short-range pairwise, complementary attractions that decorate the binding interfaces of the subunit. Each subunit has two layers of excluders and attractors. Attractor positions are arranged so that complementary attractors along a subunit-subunit interface perfectly overlap in the ground state configuration; excluders on either side of the interface are separated by exactly the cut off of their potential ($x\rsub{c}$, Eq.~\ref{eq:LJ}).  Subunits have no internal degrees of freedom -- they translate and rotate as rigid bodies.

\begin{figure}
\begin{center}
\label{fig:model}
\includegraphics{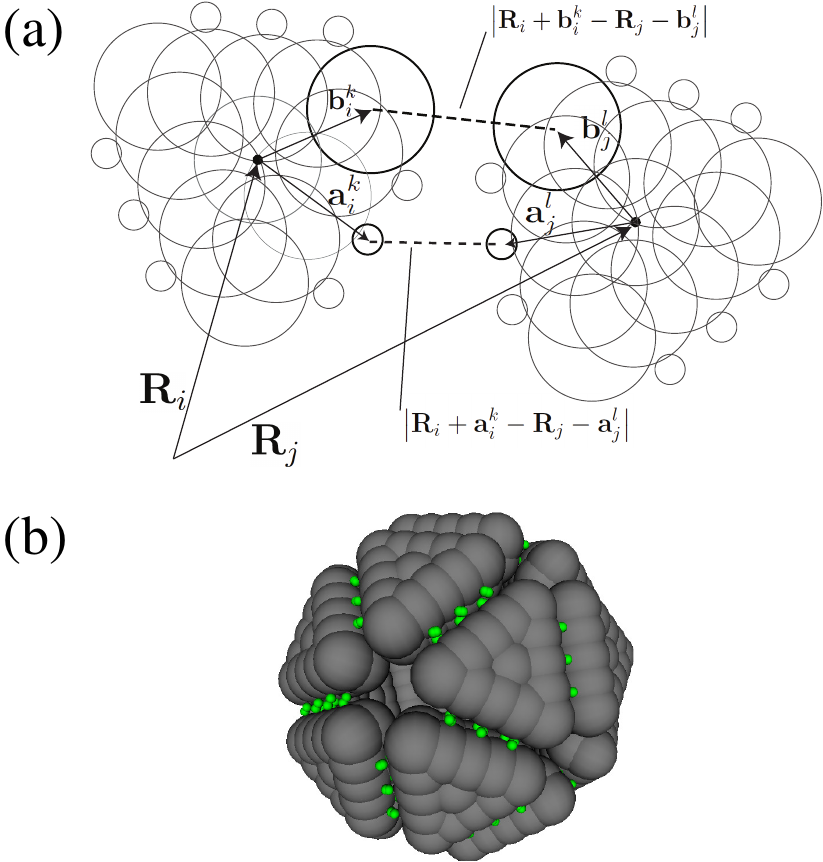}
\caption{The model capsid geometry. {\bf (a)}~Two dimensional projection of one layer of a model subunit illustrating the geometry of the capsomer-capsomer pair potential, equation (\ref{Ucc}), with a particular excluder and attractor highlighted from each subunit. The potential is the sum over all excluder-excluder and complementary attractor-attractor pairs. {\bf (b)}~An example of a well-formed model capsid from a simulation trajectory.
}
\label{fig:modelCapsid}
\end{center}
\end{figure}

The capsid subunits interact through a pairwise potential, which can be decomposed into pairwise interactions between the elemental building blocks -- the excluders and attractors. The potential of capsomer subunit $i$, $U\sub{cap,}{i}$, with position $\bm R_i$, attractor positions $\{\bm a_i\}$ and excluder positions $\{\bm b_i\}$ interacting with subunit $j$ is the sum of a repulsive potential between every pair of excluders and an attractive interaction between complementary attractors:
\begin{align}
\label{Ucc}
 & U\sub{cc}{}(\bm R_i, \{\bm a_i\}, \{\bm b_i\} ,\bm R_j, \{\bm b_j\}, \{\bm a_j\})  = \nonumber \\
  &  \sum_{k,l}^{N\sub{b}{}} \LJ{8} \left(
    \left| \bm{R}_i + \bm{b}_i^k - \bm{R}_j - \bm{b}_j^l \right|,
    \ 2^{1/4} \sigma\sub{b},
    \ \sigma\sub{b} \right) +
    \nonumber \\
    &
    \sum_{k,l}^{N\sub{a}{}} \chi_{kl} \ecc \LJ{4} \left(
    \left| \bm{R}_i + \bm{a}_i^k - \bm{R}_j - \bm{a}_j^l \right| - 2^{1/2} \sigma\sub{a},
    \ 4\sigma\sub{a},
    \ \sigma\sub{a} \right)
\end{align}
where $\ecc$ is an adjustable parameter setting the strength of the capsomer-capsomer attraction at each attractor site, $N\sub{b}{}$ and $N\sub{a}{}$ are the number of excluders and attractors respectively, $\sigma\sub{b}$ and $\sigma\sub{a}$ are the diameters of the excluders and attractors,  $\bm{b}_i^k$ ($\bm{a}_i^k)$ is the body-centered location of the $k^\mathrm{th}$ excluder (attractor) on the $i\mathrm{th}$ subunit, $\chi_{kl}$ is 1 if attractors $k$ and $l$ are overlapping in a completed capsid (Figure \ref{fig:modelCapsid}) and 0 otherwise. The diameter of attractors is set to $\sigma\sub{a}=\sigma\sub{b}/5$ for all results in this work. The function $\LJ{p}$ is defined as a truncated Lennard-Jones-like potential:
\begin{equation}
\LJ{p}(x,x_\mathrm{c},\sigma) \equiv
\left\{  \begin{array}{ll}
      \frac{1}{4}\left( \left(\frac{x}{\sigma}\right)^{-p} - \left(\frac{x}{\sigma}\right)^{-p/2} \right) & : x < x_\mathrm{c} \\
      0 & : \mathrm{otherwise}
      \end{array} \right.
      \label{eq:LJ}
\end{equation}

In our dynamical simulations, the capsomer subunits have anisotropic translational and rotational
diffusion constants, calculated as in~\cite{Hydrosub}.  The unit of time is set by the diffusion
constant of a single excluder $D$, and we define $t_0=\sigma\sub{b}^2/D$.  In these dimensionless
units, the eigenvalues of the translational and rotational diffusion tensors for capsomer subunits
are
$\{0.283, 0.283, 0.197\}$ and $\{0.1906, 0.1906, 0.0984\}$ respectively.

\section{Production rate within the steady state}
\label{app:mall}

Here we explain how we solved the kinetic equations~(\ref{equ:dtrho}) to obtain the cluster
production rate $R$ in the steady state ensemble.  As discussed in the main text, Eq.~(\ref{equ:dtrho})
with $n=M$ reduces to
\begin{equation}
\frac{\partial}{\partial t} \rho_M(t) = D \rho_1(t) [ \rho_{M-1}(t) - \rho_M(t) ] - \lambda_M \rho_M(t)
\label{equ:dtrhoM}
\end{equation}
and the production rate is
\begin{equation} R(t)=D\rho_1(t)\rho_M(t) .  \label{equ:RrhoM} \end{equation}
For completeness, we also give the equation of motion for the monomer concentration $\rho_1(t)$ within the
steady state, which is
\begin{align}
\frac{\partial}{\partial t} \rho_1(t) = &
 M R(t)
 - 2 D \rho_1(t)^2 + 2\lambda_2 \rho_2(t)   
\nonumber \\ &
+\sum_{n=2}^{M} [\lambda_{n} \rho_{n}(t) - D \rho_1(t) \rho_{n-1}(t)]
\label{equ:rho1t}
\end{align}

In the following, we work in the steady state so
we suppress all time dependence of the $\rho_n$.
Equation (\ref{equ:RrhoM}) gives $\rho_M=D\rho_1/R$ while
(\ref{equ:dtrhoM}) gives $D\rho_{M-1}= \tfrac{R}{\rho_1}(1 + \tfrac{\lambda_M}{D\rho_1})$.
The remaining $\rho_n$ may then be obtained inductively since (\ref{equ:dtrho}) reduces to
$D(\rho_{n}-\rho_{n-1})=\frac{1}{\rho_1}[\lambda_{n+1}\rho_{n+1} - \lambda_n \rho_n]$ so that $\rho_{n-1}$ is
given in terms of $\rho_m$ with $m\geq n$.
For $1\leq n \leq M-2$ we arrive at
\begin{align}
D \rho_n & = \frac{R}{\rho_1}  + \frac{1}{\rho_1}\sum_{m=n+1}^{M-1} [ \lambda_{m} \rho_{m} - \lambda_{m+1}\rho_{m+1}]
\label{equ:rhon}
\end{align}
which allows calculation of all of the $\rho_n$, in terms of $R$, $\rho_1$ and the $\lambda_n$.

A simple case is when no unbinding takes place, so that $\lambda_m=0$.  Then,
$\rho_n=\rho_1$ for all $n$, and $\rho_T = M(M+1)\rho_1/2$.  Hence the production rate for irreversible binding is
given by (\ref{equ:Rinf}).

We now turn to the problem described in the main text, where $\lambda_m=\lambda$ for $m\leq m^*$,
with $\lambda_m=0$ for $m>m^*$.
It then follows from (\ref{equ:rhon}) that
\newcommand{\lambdati}{\tilde\lambda}
\begin{equation}
\rho_n = \left\{ \begin{array}{ll}
\frac{R}{D\rho_1}, & n\geq m^*,\\
\frac{R}{D\rho_1 }S(\lambdati,m^*-n), & n<m^*. \end{array} \right.
\label{equ:rhon_specific}
\end{equation}
where $S(x,n)=(1-x^{n+1})/(1-x)$ is obtained by summing a geometrical progression and $\lambdati=\lambda/D\rho_1$.
We then sum over $n$ to obtain $\rho_\mathrm{T}$ and eliminate $R$ from the result using
\begin{equation}R=D\rho_1^2/S(\lambdati,m^*-1) \label{equ:R_rho1} \end{equation}
[which follows from (\ref{equ:rhon_specific})].
The result is
\begin{equation}
(D\rho_T/\lambda) = \frac{ (M-m^*)(M+m^*+1) + f(\lambdati,m^*) }{ 2 \lambdati S(\lambdati,m^*-1) }
\label{equ:rhoTlam}
\end{equation}
with
\begin{align}
f(\lambdati,m) = \left[ m(m+1) - 2m \frac{\partial}{\partial \lambdati} + \frac{\partial^2}{\partial \lambdati^2} \right] S(\lambdati,m)
\end{align}
[We used $\sum_{r=1}^m r x^r = x \frac{\partial}{\partial x} S(x,m)$ and similarly
$\sum_{r=2}^m r(r-1) x^r=x^2 \frac{\partial^2}{\partial x^2} S(x,m)$.]
Dimensional analysis shows that the normalised rate
$R/R^\infty$ depends only on $M$, $m^*$ and $\lambda/D\rho_\mathrm{T}$.
We therefore fix these parameters and solve (\ref{equ:rhoTlam}) numerically for $\lambdati$,
obtaining the monomer concentration $\rho_1=\lambda/(D\lambdati)$.  The rate $R$ may then be calculated from (\ref{equ:R_rho1}),
as shown in Fig.~\ref{fig:rate_analytic} and discussed in the main text.

\section{Binding free energies for the capsid model}
\label{app:freeEnergies}

\begin{figure}
\includegraphics{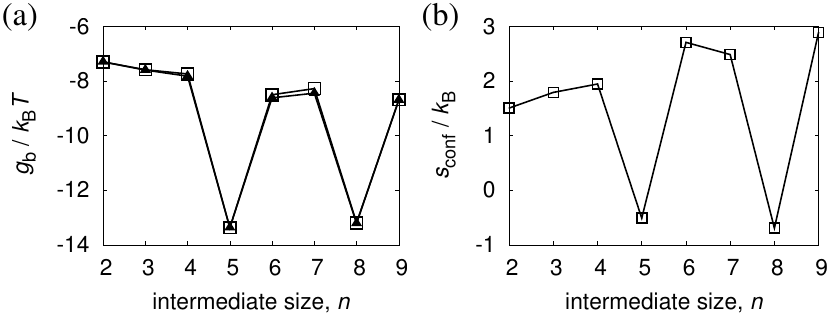}
\caption{(a)~The binding free energy $\gb$ to add an additional subunit is shown as a function of intermediate size for $\eb=4.5$. The $\blacktriangle$ symbols denote values computed from umbrella sampling simulations, while the $\Box$ symbols were calculated based on the cluster configurational entropy, as described in the text.
(b)~The change in configurational entropy, $\sconf$, computed from ground state cluster geometries is shown as a function of intermediate size. }
\label{fig:freeEnergy}
\end{figure}

\newcommand{\onkb}{/\kb}  

\renewcommand{\kb}{}
\renewcommand{\onkb}{}

We define the binding free energy (the free energy change on adding a capsomer to a cluster of size $n$) to be
\begin{equation}
\gb(n) = -\kb T \ln \left[ \frac{\rho_n}{\rho_{n-1}} \frac{\css}{\rho_1} \right]
\end{equation}
where $\rho_n$ is the concentration of clusters of size $n$ (see Sec.~\ref{sec:rate_ass})
and
$\css$ is a reference concentration (always required when quoting binding free energies).
 (We take $k_\mathrm{B}=1$ throughout this section.)
Following Ref.~\cite{Elrad2010}, by comparing the size of our capsid to the size of a satellite tobacco mosaic virus capsid, we assign $\css=8\sigma\sub{b}^{-3}$ to correspond to 1 M.

We find that the free energy of dimerization is approximately linear in $\eb$ over the range we consider:
$\gb(2) \approx -3.5 \eb - T \sbb + T \sconf(2)$ with the binding entropy penalty $\sbb\onkb=-10.7$ and
the configurational entropy change for dimerization $\sconf(2)\onkb=1.5$.
Here $\sconf(n)$ is a difference in ``configurational entropy'',
defined as $\sconf(n)=\kb \ln (\Omega_n/\Omega_{n-1})$, with $\Omega_n$ the number
of distinct ground state cluster configurations with $n$ subunits.
(In counting distinct configurations, the three edges of each capsomer are assumed to be distinguishable,
 but configurations related by global rotations are not distinct from one another.
So the number of distinct dimer configurations is $\Omega_2=(3^2/2)$ since
there are three possible binding sites on each capsomer (hence $3^2$ configurations) while the factor of 2 accounts for a rotation symmetry of the entire dimer.)
Note that the value for $\sbb$ given in Ref.~\cite{Elrad2010} contains a typo.

The binding free energy depends on the number of contacts that can be formed and the symmetry of the ground state complex. To illustrate the latter effect, we calculate $\sconf(n)$ from geometrical considerations. The approach follows Zlotnick \cite{Zlotnick1994} except that we consider all possible ground state structures. The resulting configurational entropy values for intermediates up to size $n=9$ are shown in  Fig.~\ref{fig:freeEnergy} (right), where it is evident that clusters in which every subunit has two or more bonds (e.g. $n=5$, $n=8$) have lower configurational entropy values. We discontinued the calculation at $n=10$ because the number of possible ground state structures becomes large, but the trend continues.

 The $\Box$ symbols in Fig.~\ref{fig:freeEnergy} are free energies computed using the calculated configurational entropy values, with the  interaction free energy for a single subunit-subunit interface extracted  from $g_1 = \gb(2)+T \sconf(2)$ with $\gb(2)$ extracted from the umbrella sampling results, and the interaction free energy to form two subunit-subunit interfaces extracted from  $g_2 = \gb(5)+T \sconf(5)$ .  Two separate estimates are required since $g_2 \ne 2 g_1$ because the binding entropy penalty for forming two bonds is not the same as for a single bond. The agreement between the extrapolated free energy values and those measured from umbrella sampling further illustrates the extent to which the system favors ground state configurations at equilibrium.

\end{appendix}

%


\begin{thebibliography}{49}%
\makeatletter
\providecommand \@ifxundefined [1]{%
 \@ifx{#1\undefined}
}%
\providecommand \@ifnum [1]{%
 \ifnum #1\expandafter \@firstoftwo
 \else \expandafter \@secondoftwo
 \fi
}%
\providecommand \@ifx [1]{%
 \ifx #1\expandafter \@firstoftwo
 \else \expandafter \@secondoftwo
 \fi
}%
\providecommand \natexlab [1]{#1}%
\providecommand \enquote  [1]{``#1''}%
\providecommand \bibnamefont  [1]{#1}%
\providecommand \bibfnamefont [1]{#1}%
\providecommand \citenamefont [1]{#1}%
\providecommand \href@noop [0]{\@secondoftwo}%
\providecommand \href [0]{\begingroup \@sanitize@url \@href}%
\providecommand \@href[1]{\@@startlink{#1}\@@href}%
\providecommand \@@href[1]{\endgroup#1\@@endlink}%
\providecommand \@sanitize@url [0]{\catcode `\\12\catcode `\$12\catcode
  `\&12\catcode `\#12\catcode `\^12\catcode `\_12\catcode `\%12\relax}%
\providecommand \@@startlink[1]{}%
\providecommand \@@endlink[0]{}%
\providecommand \url  [0]{\begingroup\@sanitize@url \@url }%
\providecommand \@url [1]{\endgroup\@href {#1}{\urlprefix }}%
\providecommand \urlprefix  [0]{URL }%
\providecommand \Eprint [0]{\href }%
\@ifxundefined \urlstyle {%
  \providecommand \doi  [0]{\begingroup \@sanitize@url \@doi}%
  \providecommand \@doi [1]{\endgroup \@@startlink {\doibase
  #1}doi:\discretionary {}{}{}#1\@@endlink }%
}{%
  \providecommand \doi  [0]{doi:\discretionary{}{}{}\begingroup
  \urlstyle{rm}\Url }%
}%
\providecommand \doibase [0]{http://dx.doi.org/}%
\providecommand \Doi [0]{\begingroup \@sanitize@url \@Doi }%
\providecommand \@Doi  [1]{\endgroup\@@startlink{\doibase#1}\@@Doi}%
\providecommand \@@Doi [1]{#1\@@endlink}%
\providecommand \selectlanguage [0]{\@gobble}%
\providecommand \bibinfo  [0]{\@secondoftwo}%
\providecommand \bibfield  [0]{\@secondoftwo}%
\providecommand \translation [1]{[#1]}%
\providecommand \BibitemOpen [0]{}%
\providecommand \bibitemStop [0]{}%
\providecommand \bibitemNoStop [0]{.\EOS\space}%
\providecommand \EOS [0]{\spacefactor3000\relax}%
\providecommand \BibitemShut  [1]{\csname bibitem#1\endcsname}%
\bibitem [{exp()}]{expt_virus_misc}%
  \BibitemOpen
  \href@noop {} {}\bibinfo {note} {H.~Fraenkel-Conrat and R.C. Williams, Proc.
  Natl. Acad. Sci. USA {\bf 41}, 690 (1955); A.~Klug, Phil. Trans. Royal. Soc.
  London B: Biol. Sci. {\bf 354}, 531 (1999); A.~Zlotnick~et~al., Virology {\bf
  277}, 450 (2000). J.~Sun~et al., Proc. Natl. Acad. Sci. USA {\bf 104}, 1354
  (2007).}\BibitemShut {Stop}%
\bibitem [{the()}]{theory_virus_misc}%
  \BibitemOpen
  \href@noop {} {}\bibinfo {note} {A. Zlotnick, J. Mol. Biol. {\bf 241}, 59
  (1994); B. Berger \emph{et al.}, Proc. Natl. Acad. Sci. USA {\bf 91}, 7732
  (1994); T. Chen, Z. Zhang and S.C. Glotzer, Proc. Natl. Acad. Sci. USA {\bf
  194}, 717 (2007); H.D. Nguyen, V.S. Reddy and C.L. Brooks~III, Nano Lett.
  {\bf 7}, 338 (2007).}\BibitemShut {Stop}%
\bibitem [{\citenamefont {Hagan}\ and\ \citenamefont
  {Chandler}(2006)}]{Hagan2006}%
  \BibitemOpen
  \bibfield  {author} {\bibinfo {author} {\bibfnamefont {M.~F.}\ \bibnamefont
  {Hagan}}\ and\ \bibinfo {author} {\bibfnamefont {D.}~\bibnamefont
  {Chandler}},\ }\href@noop {} {\bibfield  {journal} {\bibinfo  {journal}
  {Biophys. J.},\ }\textbf {\bibinfo {volume} {91}},\ \bibinfo {pages} {42}
  (\bibinfo {year} {2006})}\BibitemShut {NoStop}%
\bibitem [{\citenamefont {Jack}\ \emph {et~al.}(2007)\citenamefont {Jack},
  \citenamefont {Hagan},\ and\ \citenamefont {Chandler}}]{Jack2007}%
  \BibitemOpen
  \bibfield  {author} {\bibinfo {author} {\bibfnamefont {R.~L.}\ \bibnamefont
  {Jack}}, \bibinfo {author} {\bibfnamefont {M.~F.}\ \bibnamefont {Hagan}}, \
  and\ \bibinfo {author} {\bibfnamefont {D.}~\bibnamefont {Chandler}},\
  }\href@noop {} {\bibfield  {journal} {\bibinfo  {journal} {Phys. Rev. E},\
  }\textbf {\bibinfo {volume} {76}},\ \bibinfo {pages} {021119} (\bibinfo
  {year} {2007})}\BibitemShut {NoStop}%
\bibitem [{\citenamefont {Yang}\ \emph {et~al.}(2010)\citenamefont {Yang},
  \citenamefont {Meyer},\ and\ \citenamefont {Hagan}}]{Yang2010a}%
  \BibitemOpen
  \bibfield  {author} {\bibinfo {author} {\bibfnamefont {Y.}~\bibnamefont
  {Yang}}, \bibinfo {author} {\bibfnamefont {R.}~\bibnamefont {Meyer}}, \ and\
  \bibinfo {author} {\bibfnamefont {M.~F.}\ \bibnamefont {Hagan}},\ }\href@noop
  {} {\bibfield  {journal} {\bibinfo  {journal} {Phys. Rev. Lett.},\ }\textbf
  {\bibinfo {volume} {104}},\ \bibinfo {pages} {258102} (\bibinfo {year}
  {2010})}\BibitemShut {NoStop}%
\bibitem [{\citenamefont {Whitelam}(2010)}]{Whitelam2010}%
  \BibitemOpen
  \bibfield  {author} {\bibinfo {author} {\bibfnamefont {S.}~\bibnamefont
  {Whitelam}},\ }\Doi {10.1103/PhysRevLett.105.088102} {\bibfield  {journal}
  {\bibinfo  {journal} {Phys. Rev. Lett.},\ }\textbf {\bibinfo {volume}
  {105}},\ \bibinfo {pages} {088102} (\bibinfo {year} {2010})}\BibitemShut
  {NoStop}%
\bibitem [{\citenamefont {Rothemund}(2006)}]{Rothemund2006}%
  \BibitemOpen
  \bibfield  {author} {\bibinfo {author} {\bibfnamefont {P.}~\bibnamefont
  {Rothemund}},\ }\href@noop {} {\bibfield  {journal} {\bibinfo  {journal}
  {Nature},\ }\textbf {\bibinfo {volume} {440}},\ \bibinfo {pages} {297}
  (\bibinfo {year} {2006})}\BibitemShut {NoStop}%
\bibitem [{\citenamefont {Sacanna}\ \emph {et~al.}(2010)\citenamefont
  {Sacanna}, \citenamefont {Irvine}, \citenamefont {Chaikin},\ and\
  \citenamefont {Pine}}]{Sacanna2010}%
  \BibitemOpen
  \bibfield  {author} {\bibinfo {author} {\bibfnamefont {S.}~\bibnamefont
  {Sacanna}}, \bibinfo {author} {\bibfnamefont {W.~T.~M.}\ \bibnamefont
  {Irvine}}, \bibinfo {author} {\bibfnamefont {P.~M.}\ \bibnamefont {Chaikin}},
  \ and\ \bibinfo {author} {\bibfnamefont {D.~J.}\ \bibnamefont {Pine}},\
  }\href@noop {} {\bibfield  {journal} {\bibinfo  {journal} {Nature},\ }\textbf
  {\bibinfo {volume} {464}},\ \bibinfo {pages} {575} (\bibinfo {year}
  {2010})}\BibitemShut {NoStop}%
\bibitem [{\citenamefont {Caspar}\ and\ \citenamefont
  {Klug}(1962)}]{caspar-klug}%
  \BibitemOpen
  \bibfield  {author} {\bibinfo {author} {\bibfnamefont {D.~L.~D.}\
  \bibnamefont {Caspar}}\ and\ \bibinfo {author} {\bibfnamefont
  {A.}~\bibnamefont {Klug}},\ }\href@noop {} {\bibfield  {journal} {\bibinfo
  {journal} {Cold Spring Harbor Symp. Quant. Biol.},\ }\textbf {\bibinfo
  {volume} {27}},\ \bibinfo {pages} {1} (\bibinfo {year} {1962})}\BibitemShut
  {NoStop}%
\bibitem [{\citenamefont {Whitesides}\ and\ \citenamefont
  {Boncheva}(2002)}]{Whitesides2002}%
  \BibitemOpen
  \bibfield  {author} {\bibinfo {author} {\bibfnamefont {G.~M.}\ \bibnamefont
  {Whitesides}}\ and\ \bibinfo {author} {\bibfnamefont {M.}~\bibnamefont
  {Boncheva}},\ }\href@noop {} {\bibfield  {journal} {\bibinfo  {journal}
  {Proc. Natl. Acad. Sci. (USA)},\ }\textbf {\bibinfo {volume} {99}},\ \bibinfo
  {pages} {4769} (\bibinfo {year} {2002})}\BibitemShut {NoStop}%
\bibitem [{\citenamefont {Zlotnick}(2007)}]{Zlotnick2007}%
  \BibitemOpen
  \bibfield  {author} {\bibinfo {author} {\bibfnamefont {A.}~\bibnamefont
  {Zlotnick}},\ }\href@noop {} {\bibfield  {journal} {\bibinfo  {journal} {J.
  Mol. Biol.},\ }\textbf {\bibinfo {volume} {366}},\ \bibinfo {pages} {14}
  (\bibinfo {year} {2007})}\BibitemShut {NoStop}%
\bibitem [{\citenamefont {Rapaport}(2008)}]{Rapaport2008}%
  \BibitemOpen
  \bibfield  {author} {\bibinfo {author} {\bibfnamefont {D.~C.}\ \bibnamefont
  {Rapaport}},\ }\href@noop {} {\bibfield  {journal} {\bibinfo  {journal}
  {Phys. Rev. Lett.},\ }\textbf {\bibinfo {volume} {101}},\ \bibinfo {pages}
  {186101} (\bibinfo {year} {2008})}\BibitemShut {NoStop}%
\bibitem [{\citenamefont {Elrad}\ and\ \citenamefont
  {Hagan}(2008)}]{Elrad2008}%
  \BibitemOpen
  \bibfield  {author} {\bibinfo {author} {\bibfnamefont {O.~M.}\ \bibnamefont
  {Elrad}}\ and\ \bibinfo {author} {\bibfnamefont {M.~F.}\ \bibnamefont
  {Hagan}},\ }\Doi {DOI 10.1021/nl802269a} {\bibfield  {journal} {\bibinfo
  {journal} {Nano Letters},\ }\textbf {\bibinfo {volume} {8}},\ \bibinfo
  {pages} {3850} (\bibinfo {year} {2008})}\BibitemShut {NoStop}%
\bibitem [{\citenamefont {Whitelam}\ \emph {et~al.}(2009)\citenamefont
  {Whitelam}, \citenamefont {Feng}, \citenamefont {Hagan},\ and\ \citenamefont
  {Geissler}}]{Whitelam2009}%
  \BibitemOpen
  \bibfield  {author} {\bibinfo {author} {\bibfnamefont {S.}~\bibnamefont
  {Whitelam}}, \bibinfo {author} {\bibfnamefont {E.~H.}\ \bibnamefont {Feng}},
  \bibinfo {author} {\bibfnamefont {M.~F.}\ \bibnamefont {Hagan}}, \ and\
  \bibinfo {author} {\bibfnamefont {P.~L.}\ \bibnamefont {Geissler}},\
  }\href@noop {} {\bibfield  {journal} {\bibinfo  {journal} {Soft Matter},\
  }\textbf {\bibinfo {volume} {5}},\ \bibinfo {pages} {1251} (\bibinfo {year}
  {2009})}\BibitemShut {NoStop}%
\bibitem [{\citenamefont {Nguyen}\ \emph {et~al.}(2007)\citenamefont {Nguyen},
  \citenamefont {Reddy},\ and\ \citenamefont {{Brooks III}}}]{Nguyen2007}%
  \BibitemOpen
  \bibfield  {author} {\bibinfo {author} {\bibfnamefont {H.~D.}\ \bibnamefont
  {Nguyen}}, \bibinfo {author} {\bibfnamefont {V.~S.}\ \bibnamefont {Reddy}}, \
  and\ \bibinfo {author} {\bibfnamefont {C.~L.}\ \bibnamefont {{Brooks III}}},\
  }\href@noop {} {\bibfield  {journal} {\bibinfo  {journal} {Nano Lett.},\
  }\textbf {\bibinfo {volume} {7}},\ \bibinfo {pages} {338} (\bibinfo {year}
  {2007})}\BibitemShut {NoStop}%
\bibitem [{\citenamefont {Wilber}\ \emph {et~al.}(2007)\citenamefont {Wilber},
  \citenamefont {Doye}, \citenamefont {Louis}, \citenamefont {Noya},
  \citenamefont {Miller},\ and\ \citenamefont {Wong}}]{Wilber2007}%
  \BibitemOpen
  \bibfield  {author} {\bibinfo {author} {\bibfnamefont {A.~W.}\ \bibnamefont
  {Wilber}}, \bibinfo {author} {\bibfnamefont {J.~P.~K.}\ \bibnamefont {Doye}},
  \bibinfo {author} {\bibfnamefont {A.~A.}\ \bibnamefont {Louis}}, \bibinfo
  {author} {\bibfnamefont {E.~G.}\ \bibnamefont {Noya}}, \bibinfo {author}
  {\bibfnamefont {M.~A.}\ \bibnamefont {Miller}}, \ and\ \bibinfo {author}
  {\bibfnamefont {P.}~\bibnamefont {Wong}},\ }\href@noop {} {\bibfield
  {journal} {\bibinfo  {journal} {J. Chem. Phys.},\ }\textbf {\bibinfo {volume}
  {127}} (\bibinfo {year} {2007})}\BibitemShut {NoStop}%
\bibitem [{\citenamefont {Wilber}\ \emph {et~al.}(2009)\citenamefont {Wilber},
  \citenamefont {Doye},\ and\ \citenamefont {Louis}}]{Wilber2009}%
  \BibitemOpen
  \bibfield  {author} {\bibinfo {author} {\bibfnamefont {A.~W.}\ \bibnamefont
  {Wilber}}, \bibinfo {author} {\bibfnamefont {J.~P.~K.}\ \bibnamefont {Doye}},
  \ and\ \bibinfo {author} {\bibfnamefont {A.~A.}\ \bibnamefont {Louis}},\
  }\href@noop {} {\bibfield  {journal} {\bibinfo  {journal} {Journal of
  Chemical Physics},\ }\textbf {\bibinfo {volume} {131}} (\bibinfo {year}
  {2009})}\BibitemShut {NoStop}%
\bibitem [{Klo()}]{Klotsa2011}%
  \BibitemOpen
  \href@noop {} {}\bibinfo {note} {D. Klotsa and R. L. Jack, arXiv:1103.2037 to
  appear in Soft Matter (2011)}\BibitemShut {NoStop}%
\bibitem [{\citenamefont {Elrad}\ and\ \citenamefont
  {Hagan}(2010)}]{Elrad2010}%
  \BibitemOpen
  \bibfield  {author} {\bibinfo {author} {\bibfnamefont {O.}~\bibnamefont
  {Elrad}}\ and\ \bibinfo {author} {\bibfnamefont {M.~F.}\ \bibnamefont
  {Hagan}},\ }\href@noop {} {\bibfield  {journal} {\bibinfo  {journal}
  {Physical Biology},\ }\textbf {\bibinfo {volume} {7}},\ \bibinfo {pages}
  {045003} (\bibinfo {year} {2010})}\BibitemShut {NoStop}%
\bibitem [{\citenamefont {Rapaport}\ \emph {et~al.}(1999)\citenamefont
  {Rapaport}, \citenamefont {Johnson},\ and\ \citenamefont
  {Skolnick}}]{Rapaport1999}%
  \BibitemOpen
  \bibfield  {author} {\bibinfo {author} {\bibfnamefont {D.}~\bibnamefont
  {Rapaport}}, \bibinfo {author} {\bibfnamefont {J.}~\bibnamefont {Johnson}}, \
  and\ \bibinfo {author} {\bibfnamefont {J.}~\bibnamefont {Skolnick}},\
  }\href@noop {} {\bibfield  {journal} {\bibinfo  {journal} {Comput. Phys.
  Commun.},\ }\textbf {\bibinfo {volume} {121-122}},\ \bibinfo {pages} {231 }
  (\bibinfo {year} {1999})}\BibitemShut {NoStop}%
\bibitem [{\citenamefont {Rapaport}(2004)}]{Rapaport2004}%
  \BibitemOpen
  \bibfield  {author} {\bibinfo {author} {\bibfnamefont {D.}~\bibnamefont
  {Rapaport}},\ }\href@noop {} {\bibfield  {journal} {\bibinfo  {journal}
  {Phys. Rev. E},\ }\textbf {\bibinfo {volume} {70}},\ \bibinfo {pages}
  {051905} (\bibinfo {year} {2004})}\BibitemShut {NoStop}%
\bibitem [{\citenamefont {Branka}\ and\ \citenamefont
  {Heyes}(1999)}]{Branka1999}%
  \BibitemOpen
  \bibfield  {author} {\bibinfo {author} {\bibfnamefont {A.}~\bibnamefont
  {Branka}}\ and\ \bibinfo {author} {\bibfnamefont {D.}~\bibnamefont {Heyes}},\
  }\href@noop {} {\bibfield  {journal} {\bibinfo  {journal} {Phys. Rev. E},\
  }\textbf {\bibinfo {volume} {60}},\ \bibinfo {pages} {2381} (\bibinfo {year}
  {1999})}\BibitemShut {NoStop}%
\bibitem [{\citenamefont {Heyes}\ and\ \citenamefont
  {Branka}(2000)}]{Heyes2000}%
  \BibitemOpen
  \bibfield  {author} {\bibinfo {author} {\bibfnamefont {D.}~\bibnamefont
  {Heyes}}\ and\ \bibinfo {author} {\bibfnamefont {A.}~\bibnamefont {Branka}},\
  }\href@noop {} {\bibfield  {journal} {\bibinfo  {journal} {Mol. Phys.},\
  }\textbf {\bibinfo {volume} {98}},\ \bibinfo {pages} {1949} (\bibinfo {year}
  {2000})}\BibitemShut {NoStop}%
\bibitem [{\citenamefont {de~la Torre}\ and\ \citenamefont
  {Carrasco}(2002)}]{Hydrosub}%
  \BibitemOpen
  \bibfield  {author} {\bibinfo {author} {\bibfnamefont {J.~G.}\ \bibnamefont
  {de~la Torre}}\ and\ \bibinfo {author} {\bibfnamefont {B.}~\bibnamefont
  {Carrasco}},\ }\href@noop {} {\bibfield  {journal} {\bibinfo  {journal}
  {Biopolymers},\ }\textbf {\bibinfo {volume} {63}},\ \bibinfo {pages} {163}
  (\bibinfo {year} {2002})}\BibitemShut {NoStop}%
\bibitem [{Note1()}]{Note1}%
  \BibitemOpen
  \bibinfo {note} {The simulations with $\varepsilon _\protect \mathrm {b}=4.1$
  and $\varepsilon _\protect \mathrm {b}=4.3$ have not completely transitioned
  into the logarithmic growth phase at this time point. Since the nucleation
  time rises exponentially with decreasing $\varepsilon _\protect \mathrm {b}$
  \cite {Hagan2010}; the transition to logarithmic growth phase increases in a
  similar fashion. However, the variation of yield with respect to $\varepsilon
  _\protect \mathrm {b}$ is robust to this choice of observation
  time.}\BibitemShut {Stop}%
\bibitem [{Note2()}]{Note2}%
  \BibitemOpen
  \bibinfo {note} {The maximal interaction energy between subunits is $6
  \varepsilon _\protect \mathrm {b}$, so the value of $\varepsilon _\protect
  \mathrm {b}$ at optimal assembly may seem large. However, there is a
  significant entropy penalty on binding, due to the short length scale of the
  interactions and their orientation-dependence. Thus with $\varepsilon
  _\protect \mathrm {b}=4.5$ the binding free energy is approximately
  $g_\protect \mathrm {b}=-7 k_\protect \mathrm {B}T$, as discussed in
  Appendix~\ref {app:freeEnergies} and Fig.~\ref {fig:freeEnergy}.}\BibitemShut
  {Stop}%
\bibitem [{\citenamefont {Whitelam}\ and\ \citenamefont
  {Geissler}(2007)}]{Whitelam2007}%
  \BibitemOpen
  \bibfield  {author} {\bibinfo {author} {\bibfnamefont {S.}~\bibnamefont
  {Whitelam}}\ and\ \bibinfo {author} {\bibfnamefont {L.}~\bibnamefont
  {Geissler}, \bibfnamefont {Phillip}},\ }\href@noop {} {\bibfield  {journal}
  {\bibinfo  {journal} {J. Chem. Phys.},\ }\textbf {\bibinfo {volume} {127}},\
  \bibinfo {pages} {154101} (\bibinfo {year} {2007})}\BibitemShut {NoStop}%
\bibitem [{\citenamefont {Bhattacharyay}\ and\ \citenamefont
  {Troisi}(2008)}]{Bhatt2008}%
  \BibitemOpen
  \bibfield  {author} {\bibinfo {author} {\bibfnamefont {A.}~\bibnamefont
  {Bhattacharyay}}\ and\ \bibinfo {author} {\bibfnamefont {A.}~\bibnamefont
  {Troisi}},\ }\href@noop {} {\bibfield  {journal} {\bibinfo  {journal} {Chem.
  Phys. Lett.},\ }\textbf {\bibinfo {volume} {458}},\ \bibinfo {pages} {210}
  (\bibinfo {year} {2008})}\BibitemShut {NoStop}%
\bibitem [{Note3()}]{Note3}%
  \BibitemOpen
  \bibinfo {note} {If we had taken Kawasaki dynamics where MC attempted moves
  involve single particles, then the diffusion constant of clusters of
  particles would be very small when $\varepsilon _\protect \mathrm {b}$ is
  large. The resulting dynamics would not be consistent with the physical
  scenario considered here.}\BibitemShut {Stop}%
\bibitem [{\citenamefont {Maibaum}(2008)}]{Maibaum2008}%
  \BibitemOpen
  \bibfield  {author} {\bibinfo {author} {\bibfnamefont {L.}~\bibnamefont
  {Maibaum}},\ }\href@noop {} {\bibfield  {journal} {\bibinfo  {journal} {Phys.
  Rev. Lett.},\ }\textbf {\bibinfo {volume} {101}},\ \bibinfo {pages} {256102}
  (\bibinfo {year} {2008})}\BibitemShut {NoStop}%
\bibitem [{\citenamefont {Endres}\ and\ \citenamefont
  {Zlotnick}(2002)}]{Endres2002}%
  \BibitemOpen
  \bibfield  {author} {\bibinfo {author} {\bibfnamefont {D.}~\bibnamefont
  {Endres}}\ and\ \bibinfo {author} {\bibfnamefont {A.}~\bibnamefont
  {Zlotnick}},\ }\href@noop {} {\bibfield  {journal} {\bibinfo  {journal}
  {Biophys. J.},\ }\textbf {\bibinfo {volume} {83}},\ \bibinfo {pages} {1217}
  (\bibinfo {year} {2002})}\BibitemShut {NoStop}%
\bibitem [{Note4()}]{Note4}%
  \BibitemOpen
  \bibinfo {note} {We note that $Y_4^\protect \mathrm {ss}$ depends on the
  product size cutoff $n_\protect \mathrm {max}$ while $n_4(t)$ depends on the
  time $t$, so the comparison between these measurements is necessarily
  qualitative, but their respective dependencies on $n_\protect \mathrm {max}$
  and $t$ are weak and they show similar non-monotonic behaviour.}\BibitemShut
  {Stop}%
\bibitem [{Note5()}]{Note5}%
  \BibitemOpen
  \bibinfo {note} {In~\cite {Jack2007}, some of us argued that the poor
  assembly that often occurs in system with strong intercomponent bonds is
  related to the breakdown of a condition that we called local equilibration
  (our use of this term is similar in spirit to an analogous condition in
  non-equilibrium thermodynamics~\cite {noneq-thermo-book}, but the locality
  that we refer to is the space of cluster configurations, rather than in the
  spatial coordinates of the system). Here, we use the term `cluster
  equilibrium' to formulate a closely-related condition and we test the extent
  to which this condition is correlated with effective
  self-assembly.}\BibitemShut {Stop}%
\bibitem [{Note6()}]{Note6}%
  \BibitemOpen
  \bibinfo {note} {In principle, measurements of ${\protect \mathcal
  {N}_{n,\alpha }(t)}/{\protect \mathcal {N}_{n,\gamma }(t)}$ in the
  umbrella-sampled ensemble may depend on the value of $n_\protect \mathrm
  {umb}$, especially if morphologies $\alpha $ and $\gamma $ have different
  excluded volumes. In this case the cluster equilibration condition would be
  ill-defined due to cluster-cluster interactions. However, we used a range of
  values for $n_\protect \mathrm {umb}$ in our simulations and we did not
  observe any such dependence.}\BibitemShut {Stop}%
\bibitem [{\citenamefont {Becker}\ and\ \citenamefont
  {D\"oring}(1935)}]{Becker1935}%
  \BibitemOpen
  \bibfield  {author} {\bibinfo {author} {\bibfnamefont {R.}~\bibnamefont
  {Becker}}\ and\ \bibinfo {author} {\bibfnamefont {W.}~\bibnamefont
  {D\"oring}},\ }\href@noop {} {\bibfield  {journal} {\bibinfo  {journal} {Ann.
  Phys. (Leipzig)},\ }\textbf {\bibinfo {volume} {416}},\ \bibinfo {pages}
  {719} (\bibinfo {year} {1935})}\BibitemShut {NoStop}%
\bibitem [{\citenamefont {Binder}\ and\ \citenamefont
  {Stauffer}(1976)}]{Binder1976}%
  \BibitemOpen
  \bibfield  {author} {\bibinfo {author} {\bibfnamefont {K.}~\bibnamefont
  {Binder}}\ and\ \bibinfo {author} {\bibfnamefont {D.}~\bibnamefont
  {Stauffer}},\ }\href@noop {} {\bibfield  {journal} {\bibinfo  {journal} {Adv.
  Phys.},\ }\textbf {\bibinfo {volume} {25}},\ \bibinfo {pages} {343} (\bibinfo
  {year} {1976})}\BibitemShut {NoStop}%
\bibitem [{\citenamefont {Zlotnick}(2005)}]{Zlotnick2005}%
  \BibitemOpen
  \bibfield  {author} {\bibinfo {author} {\bibfnamefont {A.}~\bibnamefont
  {Zlotnick}},\ }\href@noop {} {\bibfield  {journal} {\bibinfo  {journal} {J.
  Mol. Recog.},\ }\textbf {\bibinfo {volume} {18}},\ \bibinfo {pages} {479}
  (\bibinfo {year} {2005})}\BibitemShut {NoStop}%
\bibitem [{\citenamefont {Bray}(1994)}]{Bray1994}%
  \BibitemOpen
  \bibfield  {author} {\bibinfo {author} {\bibfnamefont {A.}~\bibnamefont
  {Bray}},\ }\href@noop {} {\bibfield  {journal} {\bibinfo  {journal} {Adv
  Phys},\ }\textbf {\bibinfo {volume} {43}},\ \bibinfo {pages} {357} (\bibinfo
  {year} {1994})}\BibitemShut {NoStop}%
\bibitem [{Note7()}]{Note7}%
  \BibitemOpen
  \bibinfo {note} {In general, such rates depend on the diffusion constants and
  sizes of the relevant clusters but including such factors does not change any
  qualitative features of our results.}\BibitemShut {Stop}%
\bibitem [{\citenamefont {Hagan}\ and\ \citenamefont
  {Elrad}(2010)}]{Hagan2010}%
  \BibitemOpen
  \bibfield  {author} {\bibinfo {author} {\bibfnamefont {M.~F.}\ \bibnamefont
  {Hagan}}\ and\ \bibinfo {author} {\bibfnamefont {O.~M.}\ \bibnamefont
  {Elrad}},\ }\href@noop {} {\bibfield  {journal} {\bibinfo  {journal}
  {Biophys. J.},\ }\textbf {\bibinfo {volume} {98}},\ \bibinfo {pages} {1065}
  (\bibinfo {year} {2010})}\BibitemShut {NoStop}%
\bibitem [{\citenamefont {Zlotnick}\ \emph {et~al.}(1999)\citenamefont
  {Zlotnick}, \citenamefont {Johnson}, \citenamefont {Wingfield}, \citenamefont
  {Stahl},\ and\ \citenamefont {Endres}}]{Zlotnick1999}%
  \BibitemOpen
  \bibfield  {author} {\bibinfo {author} {\bibfnamefont {A.}~\bibnamefont
  {Zlotnick}}, \bibinfo {author} {\bibfnamefont {J.~M.}\ \bibnamefont
  {Johnson}}, \bibinfo {author} {\bibfnamefont {P.~W.}\ \bibnamefont
  {Wingfield}}, \bibinfo {author} {\bibfnamefont {S.~J.}\ \bibnamefont
  {Stahl}}, \ and\ \bibinfo {author} {\bibfnamefont {D.}~\bibnamefont
  {Endres}},\ }\href@noop {} {\bibfield  {journal} {\bibinfo  {journal}
  {Biochemistry},\ }\textbf {\bibinfo {volume} {38}},\ \bibinfo {pages} {14644}
  (\bibinfo {year} {1999})}\BibitemShut {NoStop}%
\bibitem [{\citenamefont {Moisant}\ \emph {et~al.}(2010)\citenamefont
  {Moisant}, \citenamefont {Neeman},\ and\ \citenamefont
  {Zlotnick}}]{Moisant2010}%
  \BibitemOpen
  \bibfield  {author} {\bibinfo {author} {\bibfnamefont {P.}~\bibnamefont
  {Moisant}}, \bibinfo {author} {\bibfnamefont {H.}~\bibnamefont {Neeman}}, \
  and\ \bibinfo {author} {\bibfnamefont {A.}~\bibnamefont {Zlotnick}},\
  }\href@noop {} {\bibfield  {journal} {\bibinfo  {journal} {Biophysical
  Journal},\ }\textbf {\bibinfo {volume} {99}},\ \bibinfo {pages} {1350}
  (\bibinfo {year} {2010})}\BibitemShut {NoStop}%
\bibitem [{\citenamefont {Sweeney}\ \emph {et~al.}(2008)\citenamefont
  {Sweeney}, \citenamefont {Zhang},\ and\ \citenamefont
  {Schwartz}}]{Sweeney2008}%
  \BibitemOpen
  \bibfield  {author} {\bibinfo {author} {\bibfnamefont {B.}~\bibnamefont
  {Sweeney}}, \bibinfo {author} {\bibfnamefont {T.}~\bibnamefont {Zhang}}, \
  and\ \bibinfo {author} {\bibfnamefont {R.}~\bibnamefont {Schwartz}},\
  }\href@noop {} {\bibfield  {journal} {\bibinfo  {journal} {Biophys. J.},\
  }\textbf {\bibinfo {volume} {94}},\ \bibinfo {pages} {772} (\bibinfo {year}
  {2008})}\BibitemShut {NoStop}%
\bibitem [{\citenamefont {Ceres}\ and\ \citenamefont
  {Zlotnick}(2002)}]{Ceres2002}%
  \BibitemOpen
  \bibfield  {author} {\bibinfo {author} {\bibfnamefont {P.}~\bibnamefont
  {Ceres}}\ and\ \bibinfo {author} {\bibfnamefont {A.}~\bibnamefont
  {Zlotnick}},\ }\href@noop {} {\bibfield  {journal} {\bibinfo  {journal}
  {Biochemistry},\ }\textbf {\bibinfo {volume} {41}},\ \bibinfo {pages} {11525}
  (\bibinfo {year} {2002})}\BibitemShut {NoStop}%
\bibitem [{\citenamefont {Katen}\ \emph {et~al.}(2010)\citenamefont {Katen},
  \citenamefont {Chirapu}, \citenamefont {Finn},\ and\ \citenamefont
  {Zlotnick}}]{Katen2010}%
  \BibitemOpen
  \bibfield  {author} {\bibinfo {author} {\bibfnamefont {S.~P.}\ \bibnamefont
  {Katen}}, \bibinfo {author} {\bibfnamefont {S.~R.}\ \bibnamefont {Chirapu}},
  \bibinfo {author} {\bibfnamefont {M.~G.}\ \bibnamefont {Finn}}, \ and\
  \bibinfo {author} {\bibfnamefont {A.}~\bibnamefont {Zlotnick}},\ }\href@noop
  {} {\bibfield  {journal} {\bibinfo  {journal} {Acs Chemical Biology},\
  }\textbf {\bibinfo {volume} {5}},\ \bibinfo {pages} {1125} (\bibinfo {year}
  {2010})}\BibitemShut {NoStop}%
\bibitem [{\citenamefont {Hicks}\ and\ \citenamefont
  {Henley}(2006)}]{Hicks2006}%
  \BibitemOpen
  \bibfield  {author} {\bibinfo {author} {\bibfnamefont {S.~D.}\ \bibnamefont
  {Hicks}}\ and\ \bibinfo {author} {\bibfnamefont {C.~L.}\ \bibnamefont
  {Henley}},\ }\href@noop {} {\bibfield  {journal} {\bibinfo  {journal} {Phys.
  Rev. E},\ }\textbf {\bibinfo {volume} {74}},\ \bibinfo {pages} {031912}
  (\bibinfo {year} {2006})}\BibitemShut {NoStop}%
\bibitem [{\citenamefont {Russo}\ and\ \citenamefont
  {Sciortino}(2010)}]{Russo2010}%
  \BibitemOpen
  \bibfield  {author} {\bibinfo {author} {\bibfnamefont {J.}~\bibnamefont
  {Russo}}\ and\ \bibinfo {author} {\bibfnamefont {F.}~\bibnamefont
  {Sciortino}},\ }\href@noop {} {\bibfield  {journal} {\bibinfo  {journal}
  {Phys. Rev. Lett.},\ }\textbf {\bibinfo {volume} {104}},\ \bibinfo {pages}
  {195701} (\bibinfo {year} {2010})}\BibitemShut {NoStop}%
\bibitem [{\citenamefont {Zlotnick}(1994)}]{Zlotnick1994}%
  \BibitemOpen
  \bibfield  {author} {\bibinfo {author} {\bibfnamefont {A.}~\bibnamefont
  {Zlotnick}},\ }\href@noop {} {\bibfield  {journal} {\bibinfo  {journal} {J.
  Mol. Biol.},\ }\textbf {\bibinfo {volume} {241}},\ \bibinfo {pages} {59}
  (\bibinfo {year} {1994})}\BibitemShut {NoStop}%
\bibitem [{\citenamefont {de~Groot}\ and\ \citenamefont
  {Mazur}(1984)}]{noneq-thermo-book}%
  \BibitemOpen
  \bibfield  {author} {\bibinfo {author} {\bibfnamefont {S.}~\bibnamefont
  {de~Groot}}\ and\ \bibinfo {author} {\bibfnamefont {P.}~\bibnamefont
  {Mazur}},\ }\href@noop {} {\emph {\bibinfo {title} {Non-equilibrium
  thermodynamics}}}\ (\bibinfo  {publisher} {Dover},\ \bibinfo {address}
  {Mineola NY},\ \bibinfo {year} {1984})\BibitemShut {NoStop}%
\end{thebibliography}
\end{document}